\documentclass[apj]{emulateapj}			

\newcommand{\myemail}{pzeidler@ari.uni-heidelberg.de}
\usepackage{natbib}

\shorttitle{HST WFC3/IR and ACS survey of Westerlund 2}
\shortauthors{Zeidler et al.}

\begin{document}

\title{A High-Resolution Multiband Survey of Westerlund 2 With the Hubble Space Telescope I: Is the Massive Star Cluster Double?}

\author{Peter Zeidler\altaffilmark{1,2}, Elena Sabbi\altaffilmark{2}, Antonella Nota\altaffilmark{2,3}, Eva K. Grebel\altaffilmark{1}, Monica Tosi\altaffilmark{5}, Alceste Z. Bonanos\altaffilmark{4}, Anna Pasquali\altaffilmark{1}, Carol Christian\altaffilmark{2}, Selma E. de Mink\altaffilmark{6}, \and  Leonardo Ubeda\altaffilmark{2}}

\altaffiltext{1}{Astronomisches Rechen-Institut, Zentrum f\"ur Astronomie der Universit\"at Heidelberg, M\"onchhofstr. 12-14, 69120 Heidelberg, Germany \email{\myemail}}
\altaffiltext{2}{Space Telescope Science Institute, 3700 San Martin Drive, Baltimore, MD 21218, USA}
\altaffiltext{3}{ESA, SRE Operations Devision}
\altaffiltext{4}{IAASARS, National Observatory of Athens, GR-15326 Penteli, Greece}
\altaffiltext{5}{INAF - Osservatorio Astronomico di Bologna}
\altaffiltext{6}{Astronomical Institute Anton Pannekoek, Amsterdam University, Science Park 904, 1098 XH, Amsterdam, The Netherlands}

\begin{abstract}
We present first results from a high resolution multi-band survey of the Westerlund~2 region with the \textit{Hubble} Space Telescope. Specifically, we imaged Westerlund 2 with the Advanced Camera for Surveys through the $F555W$, $F814W$, and $F658N$ filters and with the Wide Field Camera 3 in the $F125W$, $F160W$, and $F128N$ filters. We derive the first high resolution pixel-to-pixel map of the color excess $E(B-V)_g$ of the gas associated with the cluster, combining the H$\alpha$ ($F658N$) and Pa$\beta$ ($F128N$) line observations. We demonstrate that, as expected, the region is affected by significant differential reddening with a median of $E(B-V)_g=1.87$~mag. After separating the populations of cluster members and foreground contaminants using a $(F814W-F160W)$ vs. $F814W$ color-magnitude diagram, we identify a pronounced pre-main-sequence population in Westerlund~2 showing a distinct turn-on. After dereddening each star of Westerlund~2 individually in the color-magnitude diagram we find via over-plotting PARSEC isochrones that the distance is in good agreement with the literature value of $\sim4.16 \pm 0.33$~kpc. With zero-age-main-sequence fitting to two-color-diagrams, we derive a value of total to selective extinction of $R_V=3.95 \pm 0.135$. A spatial density map of the stellar content reveals that the cluster might be composed of two clumps. We estimate the same age of 0.5--2.0~Myr for both clumps. While the two clumps appear to be coeval, the northern clump shows a $\sim 20 \%$ lower stellar surface density.
\end{abstract}


\section{Introduction}
With a stellar mass of $M \ge 10^4~M_\odot$ \citep{Ascenso_07} and an estimated molecular cloud mass of $(1.7\pm0.8$ -- $7.5)\times10^5~M_\odot$ \citep[][both based on millimeter CO spectroscopy]{Furukawa_09,Dame_07}, the young, massive star cluster Westerlund 2 \citep[hereafter Wd2;][]{Westerlund_61} is one of the most massive young star clusters known in the Milky Way (MW). It is embedded in the \ion{H}{2} region \object{RCW~49} \citep{Rodgers_60} and located in the Carina-Sagittarius spiral arm $(\alpha,\delta)=(10^h23^m58^s.1,-57^\circ45'49'')$(J2000), $(l,b)=(284.3^\circ,-0.34^\circ)$. \citet{Moffat_91} suggest that Wd2 contains more than 80 O-type stars.

Very few young ($<5$~Myr) and massive ($>10^4~M_\odot$) star clusters are known in the MW, but they are frequently observed in the \object{Magellanic Clouds} \citep[e.g., ][]{Gascoigne_52,Clark_05} as well as in the more distant Universe, especially in disk and starburst galaxies \citep[e.g.,][]{Hodge_61}. Therefore, Wd2 is a perfect target to study the star formation process and feedback of the gas in the presence of massive stars as well as the possible triggering of star formation in the surrounding molecular cloud \citep[suggested to occur in \object{RCW~49} by][using \textit{Spitzer} mid-IR images]{Whitney_04,Churchwell_04}. In fact, many globular clusters are less massive than Wd2 \citep[see, e.g.,][]{Misgeld_11}.

The Wd2 cluster has been widely discussed in the literature in the last decade, yet its physical properties are poorly known. There is a considerable disagreement about the distance of Wd2, ranging from values of 2.8~kpc \citep{Rauw_07,Ascenso_07,Carraro_13}, 4.16~kpc \citep{Alvarez_13}, 5.7~kpc \citep{Piatti_98}, 6.4~kpc \citep{Carraro_04} to even 8~kpc \citep{Rauw_07,Rauw_11}. A consensus has not been reached on the age of Wd2 either, although there seems to be general agreement that the cluster as a whole is younger than 3~Myr and that the core may be even younger than 2~Myr \citep{Ascenso_07,Carraro_13}. This young age combined with the cluster half-light radius of about 2.4~pc and its total mass make it likely that Wd2 is younger than half of its typical crossing time \citep[$\sim 5.5$~Myr][]{Gieles_11}. Therefore, the physical conditions of Wd2 are very close to its initial conditions.

Over the last years a series of spectroscopic observations of the most massive stars was performed. \citet{MSP_91} used $UBV$ photometry and low resolution spectroscopy in order to classify six O-stars within Wd2 as O6--7~V objects and to obtain a distance of $d=7.9^{+1.2}_{-1.0}$~kpc, while \citet{Piatti_98} reanalyzed a subset of O stars from \citep{MSP_91} to determine a distance of $d=5.7 \pm 0.3$~kpc. Two snapshot spectra of \object{WR20a} were taken and analyzed by \citet{Rauw_04}. They found that \object{WR20a} is a double-lined binary with a preferred orbital period of $3.675 \pm 0.030$~days with component minimum masses of $68.8 \pm 3.8~M_\odot$ and $70.7 \pm 4.0~M_\odot$. They suggest a spectral type of WN6ha for both components. \citet{Bonanos_04} discovered it was also an eclipsing binary and refined those values to even more extreme masses of $ 83.0 \pm 5.0~M_\odot$ and $82.0 \pm 5.0~M_\odot$ since they also measured the inclination. Spectroscopic observations of \citet{Alvarez_13,Carraro_13} of the most massive stars show that Wd2 has a total to selective extinction in the visual ($R_V=A_V/E(B-V)$) of about 3.77--3.85, therefore higher than the Galactic mean \citep{Mathis_81} of $R_V=3.1$.

The proximity to our solar system makes Wd2 a good target to resolve and characterize the individual components of its stellar content. Still, Wd2 has not been extensively studied. We attempted to fill the observational gap by designing a high resolution multi-band survey of the Westerlund~2 region with the \textit{Hubble} Space Telescope. In this paper, we present new high resolution photometry, obtained with the Advanced Camera for Surveys \citep[hereafter ACS;][]{ACS} and the infrared channel of the Wide Field Camera 3 \citep[hereafter WFC3/IR;][]{WFC3} of the \textit{Hubble} Space Telescope (HST). Combining four wide band filters ($F555W$, $F814W$, $F125W$, $F160W$) we aim to constrain the distance and age of the cluster and to characterize its low-mass stellar content. Combining two narrow band filters ($F658N$, $F128N$) we can derive a pixel-to-pixel reddening map \citep{Pang_11} with the H$\alpha$ to Pa$\beta$ ratio to get the $E(B-V)$ color excess of the gas content in Wd2. This map helps us to deredden the stellar content while taking into account differential reddening.

In Section \ref{sec:observations} we describe the HST observations of Wd2, and the data reduction, based on the HST pipelines \citep{Dolphin_00,Fruchter_10}. In Section \ref{sec:catalog_creation} we describe the creation of our photometric catalog as well as the cleaning process from artifacts and spurious objects. In Section \ref{sec:reddening_map} we create the high resolution two-dimensional map of the color excess. In Section \ref{sec:stellar_reddening} we describe the transformation of the gas excess to a stellar excess. In Section \ref{sec:CMD} we describe the morphology and limitations of our color-magnitude diagram. In Section \ref{sec:physical_parameters} we determine the physical parameters of Westerlund 2 such as distance, age, and the total-to-selective extinction towards Wd2. In Section \ref{sec:two_clumps}, we present the unusual morphology of Wd2 that suggests that Wd2 might actually be a double cluster. In Section \ref{sec:summary} we summarize findings and conclusions.

\section{HST observations and data reduction}
\label{sec:observations}
\subsection{Observations}
Our observations of Wd2 were performed with the HST during Cycle 20 using the ACS \citep{ACS} and the Infrared Channel of the WFC3/IR \citep{WFC3}. Six orbits were granted and the science images were taken on 2013 September 2--8 (ID: 13038, PI: A. Nota).

We acquired data in six different filters: four wide band filters and two narrow band filters. We imaged Wd2 with ACS (resolution $0.049''$/px) through the $F555W$ and $F814W$ filters, corresponding to the $V$ and $I$-bands, as well as the $F658N$ filter, centered on the H$\alpha$ line. We also observed the cluster with WFC3/IR (resolution $0.135''$/px) and the $F125W$ and $F160W$ filters matching the $J$ and $H$-band, as well as the $F128N$ filter corresponding to the Pa$\beta$ line. In addition, short exposures were acquired in the $F555W$ and $F814W$ bands (hereafter denoted as $F555W$(s) and $F814W$(s)) to reduce saturation due to O-type stars in the very bright cluster center and across the field of view.

\begin{figure}[htb]
\resizebox{\hsize}{!}{\includegraphics{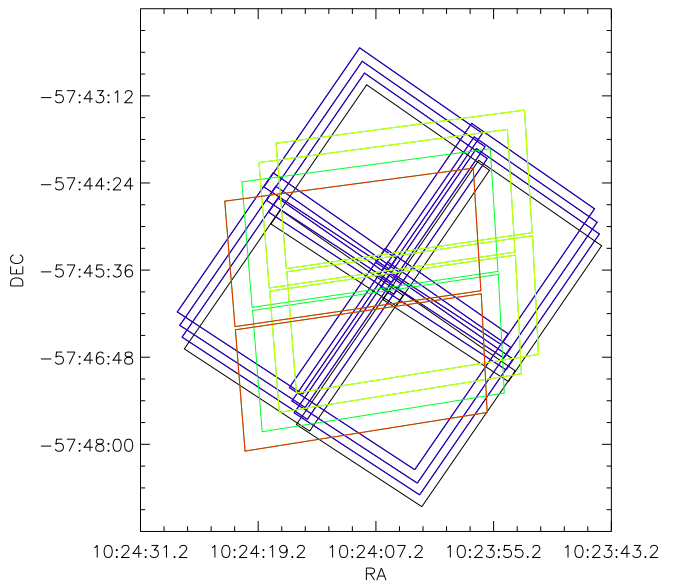}}
\caption{The field of view for the different exposures taken with the WFC3/IR and ACS filters. The blue and black fields together represent the $F125W$ and $F160W$ WFC3/IR exposures, while the blue fields alone represent the $F128N$ WFC3/IR dithered images. The yellow, green, and red fields together represent the ACS $F814W$ and $F555W$ long exposures, while the yellow fields represent the $F658N$ dither positions. The red field alone represents the short exposures for ACS $F814W$ and $F555W$.}
\label{fig:survey_area}
\end{figure}

We used one ACS pointing to cover Wd2 and a dither pattern with four positions for the long-exposures (350~s each) in the broad band filters, subsequently replaced by a two-position dithering for the $F658N$ filter (for a total exposure time of 1400~s). The short exposures were only observed at one position, each with an exposure time of 3~s. For the WFC3/IR observations mosaics of $2\times2$ tiles were required since the WFC3/IR has a smaller field-of-view (FOV) than the ACS (FOV$_{\rm{ACS}}=202 \times 202 $~arcsec and FOV$_{\rm{WFC3}}= 123 \times 136$~arcsec). Each wide-band tile consists of four dither positions with a total exposure time of 947~s ($3\times$250~s and $1\times$200~s). The narrow band tiles consist of three dither positions with a total exposure time of 748~s (250~s each). The log of the observations is reported in Tab. \ref{tab:oberservation_overview} together with the filters and exposure times used. An overview of the spatial positions of all images as well as their alignment is shown in Fig. \ref{fig:survey_area}.

\begin{deluxetable*}{lllrrrrr}
\tablecaption{Overview of the observations \label{tab:oberservation_overview}}
\tablewidth{0pt}
\tablehead{
\multicolumn{1}{c}{\textsc{Dataset}} & \multicolumn{1}{c}{\textsc{Filter}} &\multicolumn{1}{c}{\textsc{Camera}} & \multicolumn{1}{c}{\textsc{RA(J2000)}} & \multicolumn{1}{c}{\textsc{Dec(J2000)}} & \multicolumn{1}{c}{\textsc{Exp. Time [s]}} & \multicolumn{1}{c}{\textsc{\# Exp.}} &   \multicolumn{1}{c}{\textsc{Date}}
}
\tablecolumns{8}
\startdata

IBYA01010	& $F555W$		&	ACS   &	10:24:08.989  &	 -57:46:20.95	&	1400	& 4&	2013-09-02\\
IBYA01010	& $F555W$(s)	&	ACS   &	10:24:08.989  &	 -57:46:20.95	&	3		& 1&	2013-09-02\\
JBYA01020	& $F658N$		&	ACS   &	10:24:03.756  & -57:45:33.02	&	1400	& 2&	2013-09-02\\
JBYA01030	& $F814W$		&	ACS   &	10:24:08.989  &	-57:46:20.95   	&	1400	& 4&	2013-09-02\\
JBYA01030	& $F814W$(s)	&	ACS   &	10:24:08.989  &	-57:46:20.95   	&	3		& 4&	2013-09-02\\
IBYA08010   & $F125W$		&	WFC3/IR &  10:24:07.294 &	-57:44:05.26	&	947 & 4 & 2013-09-02 \\
IBYA09010	& $F125W$		&	WFC3/IR &  10:24:16.116 &	-57:45:48.56	&	947& 4&	2013-09-02\\
IBYA10010	& $F125W$		&	WFC3/IR & 	10:24:55.869 &	-57:45:07.69	&	947& 4&	2013-09-03\\
IBYA11010	& $F125W$		&	WFC3/IR & 	10:24:04.686 &	-57:46:51.05	&	947		& 4&	2013-09-08\\
IBYA08030	& $F128N$		&	WFC3/IR &  10:24:07.294 &	-57:44:05.26	&   748 & 3&	2013-09-02\\
IBYA09030	& $F128N$		&	WFC3/IR & 	10:24:16.116 &	-57:45:48.56	&	748& 3&	2013-09-02\\
IBYA10030	& $F128N$		&	WFC3/IR & 	10:24:55.869 &	-57:45:07.69	&	748& 3&	2013-09-03\\
IBYA11030	& $F128N$		&	WFC3/IR & 	10:24:04.686 &	-57:46:51.05	&	748		& 3&	2013-09-08\\
IBYA08020	& $F160W$		&	WFC3/IR &  10:24:07.294 &	-57:44:05.26	&	947 & 4 & 2013-09-02 \\
IBYA09020	& $F160W$		&	WFC3/IR &  10:24:16.116 &	-57:45:48.56	&	947& 4&	2013-09-02\\
IBYA10020	& $F160W$		&	WFC3/IR & 	10:24:55.869 &	-57:45:07.69	&	947& 4&	2013-09-03\\
IBYA11020	& $F160W$		&	WFC3/IR & 	10:24:04.686 &	-57:46:51.05	&	947		& 4&	2013-09-08\\

\enddata
\tablecomments{This table summarizes the observations of Wd2 taken with the ACS and WFC3/IR cameras aboard HST. Given are the identifiers of the datasets, the filter, camera, as well as the central coordinates of each dither position. The exposure time is the total exposure time of all dither positions combined. We also give the observation date for each image.}
\end{deluxetable*}

\subsection{Data reduction}
The observed data were first processed with the internal pipeline \citep{ACS,WFC3} (from now on referred to as raw data), which delivers pipeline flat-fielded files (\texttt{flt.fits} for WFC3) or charge-transfer efficiency (CTE) corrected images (\texttt{flc.fits} for ACS).

The reduction of the raw data consists of multiple steps performed by AstroDrizzle \citep{Fruchter_10}, the follow-up tool of MultiDrizzle in the HST pipeline. The major goal of AstroDrizzle is to reconstruct and recover the spatial information of a set of images while preserving the signal-to-noise ratio (SNR). In AstroDrizzle this is done by the method formally known as variable-pixel linear reconstruction (commonly known as Drizzle) developed by Andy Fruchter and Richard Hook \citep{Fruchter_97}. A detailed description of the package can be found in the manual of DrizzlePac (\url{http://drizzlepac.stsci.edu}).

AstroDrizzle performs its tasks on pipeline flat-fielded files or CTE corrected images. The general concept of AstroDrizzle is to partly recover the high frequency information smeared out by the detector pixel response function by combining the sub-pixel dithered images. This algorithm gives us the chance of recovering spatial information of the under-sampled WFC3/IR images to reconstruct a final combined image with a higher resolution. In the case of multiple dither-positions the final combined (drizzled) image is free of cosmic rays and detector defects. At the end all images are aligned to the same reference frame, provided by the world coordinate system (WCS) information in the header of one of the images (we chose the  F814W filter as reference).

Since the instruments (ACS and WFC3) and the pipeline data products are substantially different we treat the related images separately.

\subsubsection{The ACS data}
During the whole process the short exposures for the $F555W$ and the $F814W$ filters are treated individually and are labeled as $F555W$(s) and $F814W$(s), since AstroDrizzle cannot effectively combine images of very different exposure times.

For all ACS images we used a \texttt{pxfrac}=0.9 for the drizzling process meaning that each pixel is shrunk by 10\% in size before being mapped onto the output grid. This value was chosen so that the size of the point-spread-function (PSF) of the drizzled image reaches a minimum while mapping all pixels properly to the output grid (see Sect. \ref{subsec:quality}). The output pixel grid has the same pixel size as the original image ($0.049$~arcsec/pixel). Since there is only one dither position available for the short exposures (see Fig. \ref{fig:survey_area}), the cosmic-ray and detector defect removal could not be accomplished. Nevertheless we drizzled those images to correct them for geometric distortion and to align them properly to the reference frame. The short exposures are only taken to recover the photometry of sources that are saturated in the long exposures.

\subsubsection{The WFC3/IR data}
\label{subsec:WFC3_data}
The PSF of the IR data is under-sampled. Therefore, we reduced the pixel size of the output grid to recover as much spatial information as possible for the determination of the optimal pixel size. If the pixel size in combination with the pixel fraction is shrunk too much, there is not enough information left for mapping the pixels properly onto the output grid without generating "holes" of empty pixels. Taking into account the weighted maps while representing the contribution of each pixel to the final image and the final full width at half maximum (FWHM) of the PSFs of some representative stars, we chose a final pixel size of $0.098$~arcsec/px with a \texttt{pxfrac}=0.8. The data would have allowed us to reduce the pixel size a little more, but we found it convenient to have a pixel size twice that of the ACS. In addition, we mosaicked the four pointings into a single image.

In the end we had five mosaics for the ACS data set, one for the narrow-band and two (short and long exposures) for the wide-band filters. For the WFC3/IR filters we had three mosaics in total, one for each filter.

\subsection{Quality check of the final images}
\label{subsec:quality}

In order to check the quality of the drizzled ACS images, we tested the stability of the weighted maps of the final products. In order to determine the variations in the weighted maps, 28 boxes (20 for the $F658N$ filter) of $100\times100$~pixel and 16 boxes of $250\times250$~pixel were selected across the image. In each box the standard deviation of the median was determined using the IRAF\footnote{IRAF is contributed by the National Optical Astronomy Observatory, which is operated by the Association for Research in Astronomy, Inc., under cooperative agreement with the National Science Foundation.} task \texttt{IMEXAMINE}. This ratio should be below 0.2 following the DrizzlePac manual, meaning that the deviation of the weights of the different pixels should be less than 20\%. This test could not be performed for the IR data due to the different readout method. The results are shown in Table \ref{tab:weightend_map_summary}.

In addition, we determined the FWHM of a sample of 25 carefully selected stars across each field with the IRAF task \texttt{IMEXAMINE}. We chose medium bright, isolated, non-saturated stars so that the PSFs are not contaminated by close neighbors or remaining detector artifacts. For the PSF determination, all ACS images where divided into four quadrants. The mean FWHM of the 25 selected stars in each filter is listed in Tab. \ref{tab:mean_FWHM}. The FWHM range over the whole chip is given in the ACS manual \citep{ACS} with the FWHM of $F555M=2.12$--$2.49$~pixel as reference. The FWHM is quite constant between the different ACS filters and even the WFC3/IR filters. Our determined values are within the PSF range in the $F555M$ filter, meaning that quality and, therefore, the drizzle parameters are fine.

\begin{deluxetable}{lrrrrr}
\tablecaption{Summary of the weighted maps of the AstroDrizzle pipeline \label{tab:weightend_map_summary}}
\tablewidth{0pt}
\tablehead{
\multicolumn{1}{c}{\textsc{Filter}} &\multicolumn{1}{c}{\textsc{\# Boxes}} & \multicolumn{1}{c}{\textsc{Boxsize} [px]} &  \multicolumn{1}{c}{median} & \multicolumn{1}{c}{$\sigma$} & \multicolumn{1}{c}{$\frac{\sigma}{\mathrm{median}}$}
}
\tablecolumns{6}
\startdata
$F555W$ &$28$ & $100 \times 100$  & $ 1318 $ & $ 147.7 $ & $0.1147$  \\
$F555W$ &$16$ & $250 \times 250$  & $ 1325 $ & $ 142.2 $ & $0.1099$  \\[0.1cm]
$F658N$ &$20$ & $100 \times 100$  & $ 1336 $ & $ 208.5 $ & $0.1618$  \\
$F658N$ &$16$ & $250 \times 250$  & $ 1339 $ & $ 206.7 $ & $0.1599$  \\[0.1cm]
$F814W$ &$28$ & $100 \times 100$  & $ 1317 $ & $ 147.3 $ & $0.1145$ \\
$F814W$ &$16$ & $250 \times 250$  & $ 1322 $ & $ 147.7 $ & $0.1137$ \\
\enddata
\tablecomments{In this table we give a summary of the tests for the weighted maps. The first column shows the corresponding filter. Column 2 gives the number of boxes per image. Column 3 lists the boxsizes for calculating the median and the standard deviation of the pixel values. Column 4 and 5 show the  median and the standard deviation of the pixel values, respectively. Column 6 lists the ratio of the standard deviation and the median.}
\end{deluxetable}

\begin{deluxetable*}{lrrrr}
\tablecaption{The mean PSF FWHM in the different filters \label{tab:mean_FWHM}}
\tablewidth{0pt}
\tablehead{
\multicolumn{1}{c}{\textsc{Filter}}  &  \multicolumn{1}{c}{FWHM [px]} & \multicolumn{1}{c}{$\sigma_{\textrm{\tiny FWHM}}$ [px]} & \multicolumn{1}{c}{FWHM [arcsec]} & \multicolumn{1}{c}{$\sigma_{\textrm{\tiny FWHM}}$ [arcsec]}}

\tablecolumns{5}
\startdata
$F555W$ &$2.154$ & $0.1044$  & $0.1055$ & $0.0125$	\\
$F658N$ &$2.034$ & $0.3848$  & $0.0997$ & $0.0187$	\\
$F814W$ &$2.011$ & $0.2304$  & $0.0986$ & $0.0113$	\\
$F125W$ &$2.188$ & $0.2145$  & $0.2144$ & $0.0210$	\\
$F128N$ &$2.410$ & $0.2362$  & $0.2362$ & $0.0231$	\\
$F160W$ &$2.242$ & $0.2197$  & $0.2197$ & $0.0215$	\\
\enddata
\tablecomments{The mean FWHM of the PSF of 25 selected stars in each filter. The first column shows the corresponding filter. Column 2 and 3 show the FWHM and the standard deviation in pixels and column 4 and 5 show the FWHM and the standard deviation in arcsec.}
\end{deluxetable*}

\begin{figure*}[htb]
\resizebox{\hsize}{!}{\includegraphics{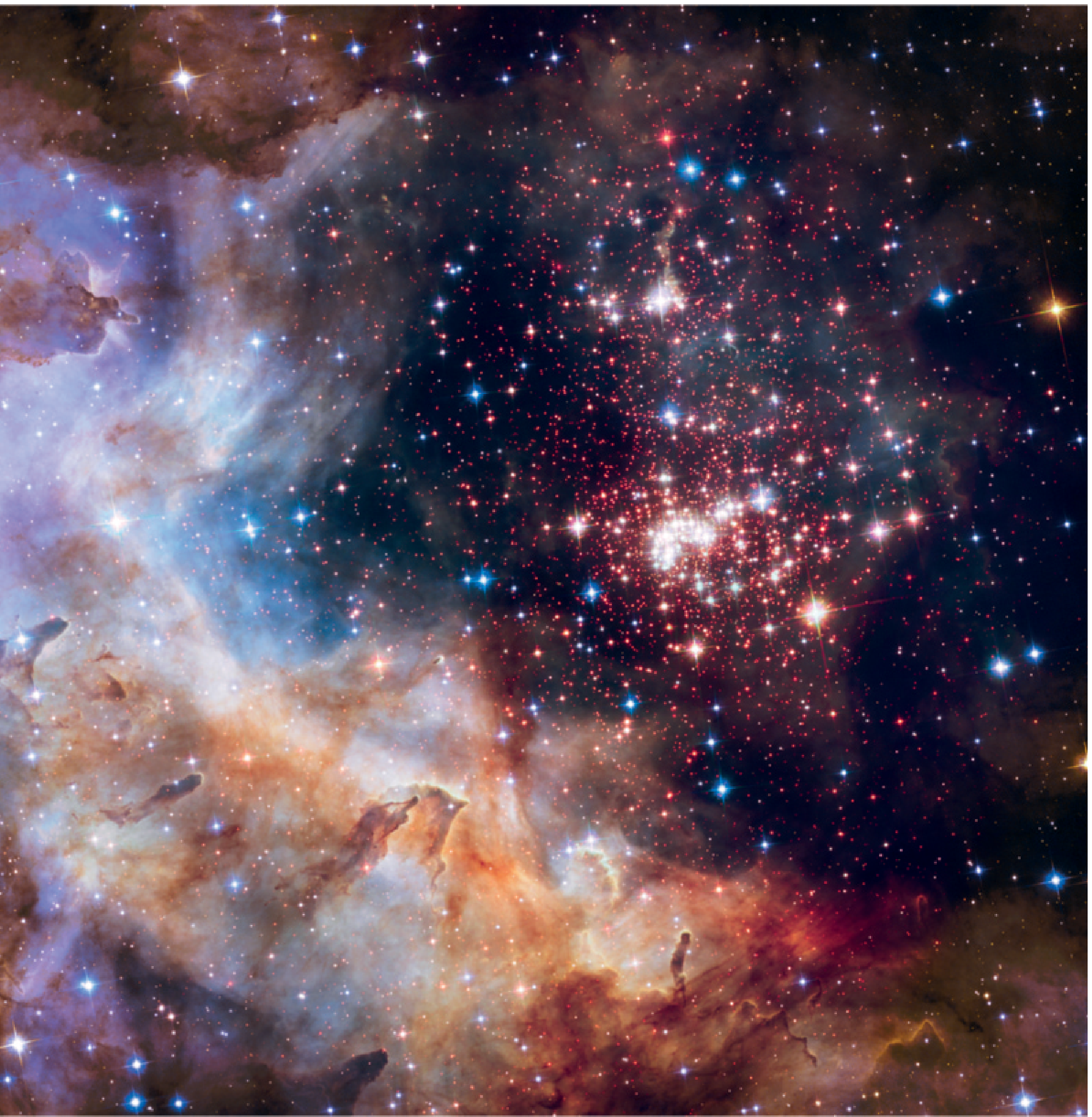}}
\caption{Color composite image of the HST ACS and WFC3/IR data of Wd2, including the $F125W$ (red), $F814W$ (green), and $F555W$ (blue) filters. North is up, East to the left. The FOV is $\sim$4~arcmin$\times$4~arcmin. This image was chosen to be the official Hubble 25th anniversary image\footnote{\url{http://hubblesite.org/newscenter/archive/releases/2015/12/image/a/}}. Credit: NASA, ESA, the Hubble Heritage Team (STScI/AURA), A. Nota (ESA/STScI), and the Westerlund 2 Science Team.}
\label{fig:RGB_JIV}
\end{figure*}

In Fig. \ref{fig:RGB_JIV} we show a color composite image of the Wd2 region, where the $F125W$, $F814W$, and $F555W$ filters are represented in red, green, and blue, respectively.

\section{The creation of the photometric catalog}
\label{sec:catalog_creation}

The source detection and photometry was performed with the DOLPHOT\footnote{\url{http://americano.dolphinsim.com/dolphot/}} package including the extensions for ACS and WFC3/IR written by Andrew Dolphin. DOLPHOT is a modified version of HSTphot \citep{Dolphin_00}.

\subsection{Source detection and photometry}
\label{subsec:source_detection}
For the source detection and photometry we adopted individual parameters for each filter. We considered the size of the PSFs, the different saturation limits, and the nebulosity conditions (due to the very different level of gas and stellar density). The main effort was to detect all sources and yet to avoid as much contamination by spurious objects as possible in order to make the cleaning procedure later-on easier. Spurious detections may happen in particular in regions with varying background brightness due to the varying gas and dust density in the \ion{H}{2} region and stellar crowding.

DOLPHOT uses the drizzled images for the source detection and works on images (flcs and flts) with geometric-distortion-corrected headers (performed by AstroDrizzle) (from now on called processed images). To perform the source detection several steps have to be applied in a given order: \texttt{acsmask} (\texttt{wfc3mask}) to mask the processed images for detector artifacts and other defects using the weighted images, \texttt{splitgroups} to create single chip and single layered fits files, and \texttt{calcsky} to create a \texttt{skymap} for all processed images.

After those preparatory steps the actual \texttt{dolphot} routine was used to create a point-source catalog using the PSF and pixel area maps from an analytic model of ACS and WFC3/IR which is provided on the DOLPHOT webpage. The photometric run was performed in one step using the reference frame ($F814W$) to obtain the common astrometry and in order to improve the photometric accuracy. The output of DOLPHOT is one catalog for each filter and exposure, so in total 8 catalogs.

\subsection{Improving the quality of the Photometric Catalogs}
\label{subsec:cleaning}

Initially, the photometric catalogs contain a number of spurious objects, which must be removed. We "cleaned" the photometric catalogs by applying appropriately chosen trimming parameters for each of the filters. The trimming values for the WFC3/IR and ACS instruments are given in the DOLPHOT manual. In order to ensure that as many spurious objects as possible were deleted, still preserving all 'real' objects, especially in the nebular regions and the dense cluster center we had to find the optimal parameter set for DOLPHOT. These values were adjusted by visual inspection of the observations of the removed and the kept objects.

\subsubsection{Object Type}
\begin{figure}[htb]
\resizebox{\hsize}{!}{\includegraphics{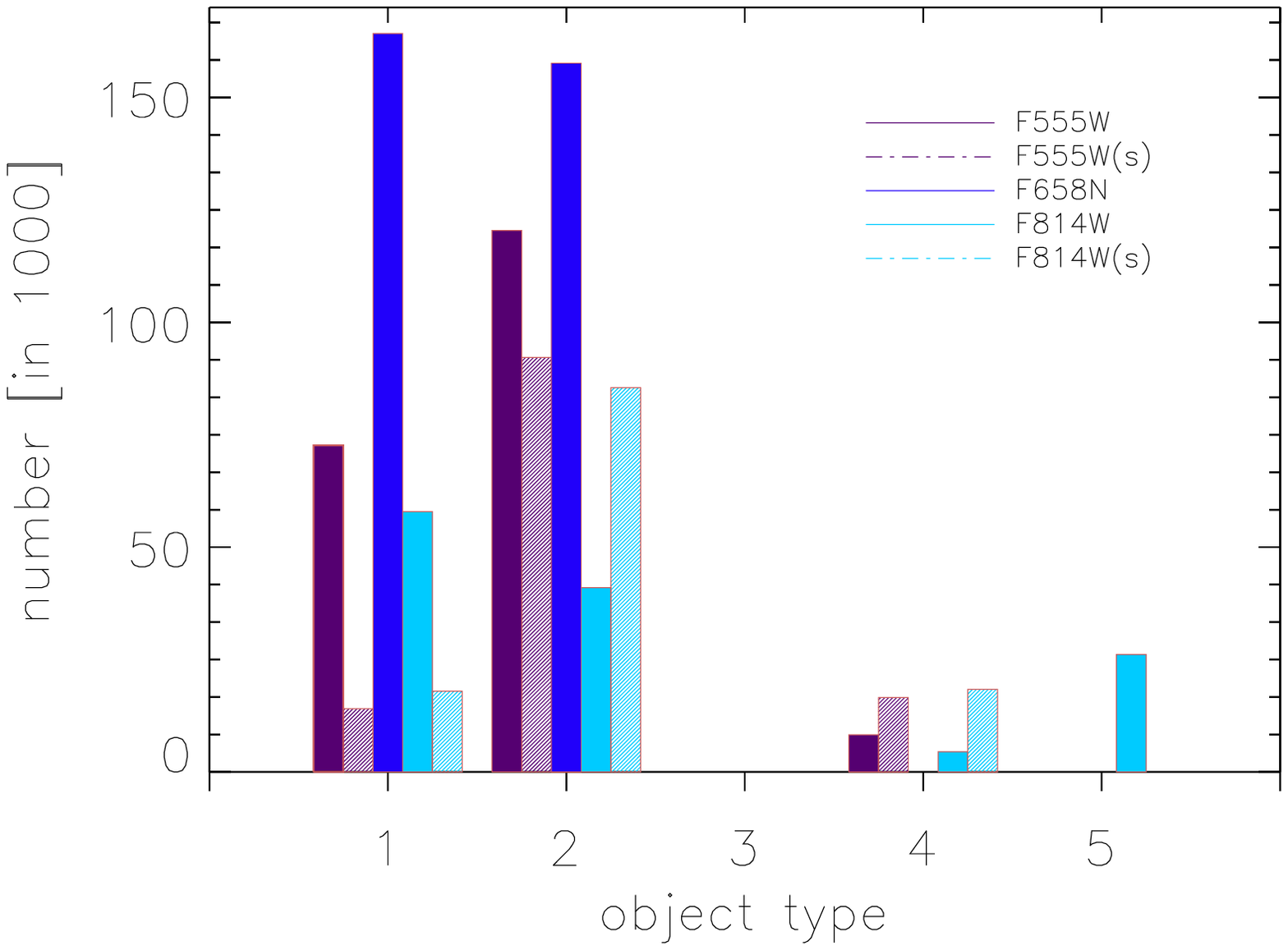}}
\caption{The total number of stars for different object types per ACS filter in the catalogs created by DOLPHOT. We decided to keep only those objects with OT=1 to reduce the number of spurious detections.}
\label{fig:object_type}
\end{figure}

The objects are classified as different types (OT) by DOLPHOT: good star (1), star too faint for PSF determination (2), elongated object (3), object too "sharp" (4) (see Sect. \ref{subsubsec:sharpness}), extended object (5). We only kept the objects with OT=1. For the WFC3/IR filters we did not select by object type in order not to lose any objects. The distribution of the OTs for the ACS filters can be seen in Fig \ref{fig:object_type}.

\subsubsection{Magnitude error}
We restricted the catalogs to contain only those stars with magnitude errors above a given threshold, which depends on the filter. The cuts for the different filters are listed in column 2 of Tab. \ref{tab:cleaning_thresholds}.

\subsubsection{Sharpness}
\label{subsubsec:sharpness}
\begin{figure*}[htb]
\resizebox{\hsize}{!}{\includegraphics{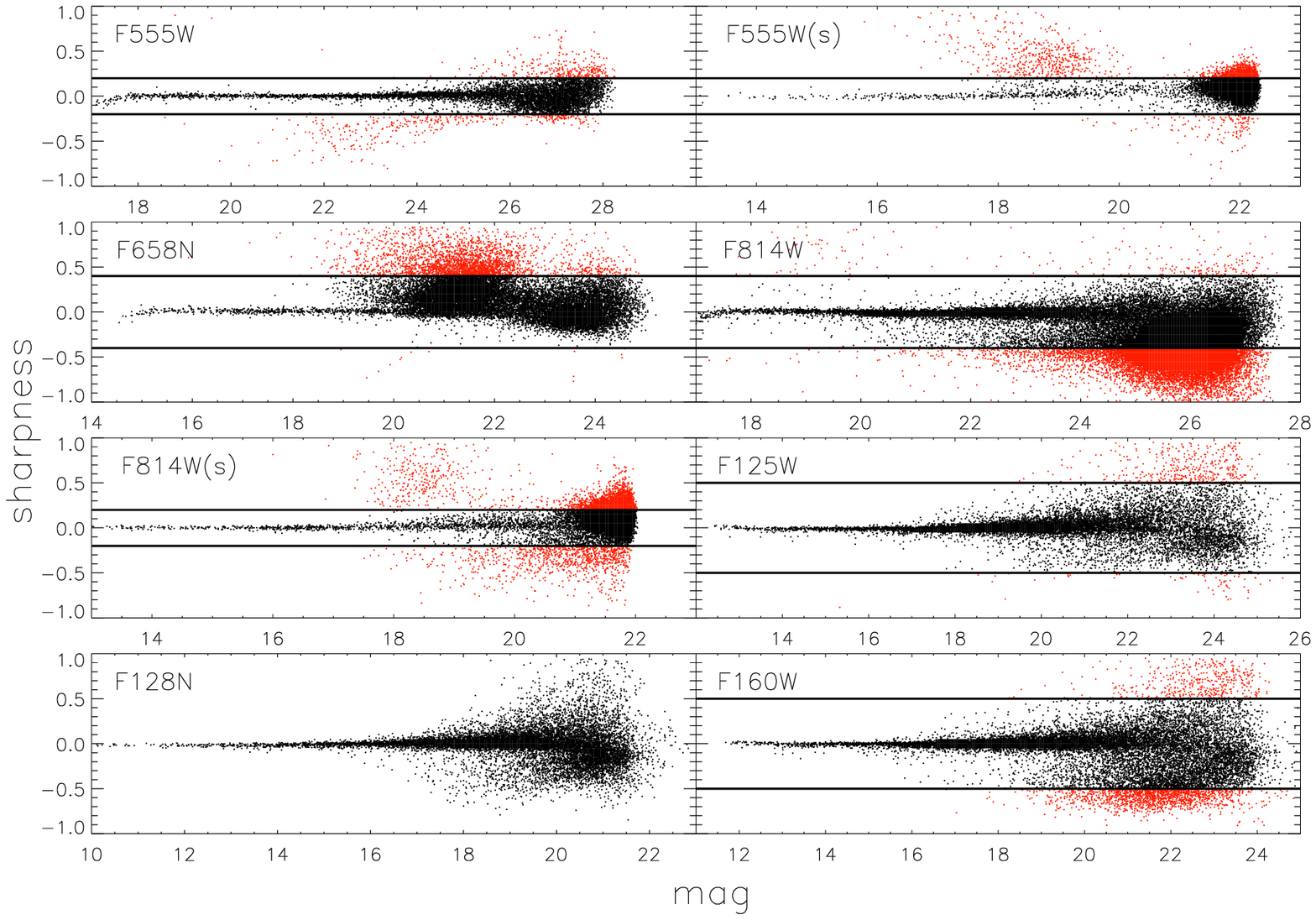}}
\caption{The distribution of sharpness, which is a measure of how peaked the source is compared to a stellar PSF as function of magnitude for each filter. The black dots represent the data retained.}
\label{fig:sharpness}
\end{figure*}

The "sharpness" parameter allows us to assess how well the PSF of a star is fitted. For a perfectly-fitted star the value is zero, positive for a point-source that is too peaked as compared to a typical stellar PSF such as a cosmic ray, and negative for a point source that is too broad in its light profile and that might be a background galaxy. Typical values range from $-0.3$ to $+0.3$ for un uncrowded region. The values chosen as our selection criteria are given in column 3 of Tab. \ref{tab:cleaning_thresholds}. For the $F128N$ filter a cut was impossible without losing "real" objects. The distribution of the sharpness as a function of magnitude can be seen in Fig. \ref{fig:sharpness}.

\subsubsection{Crowding}
\label{subsec:crowding}
\begin{figure*}[htb]
\resizebox{\hsize}{!}{\includegraphics{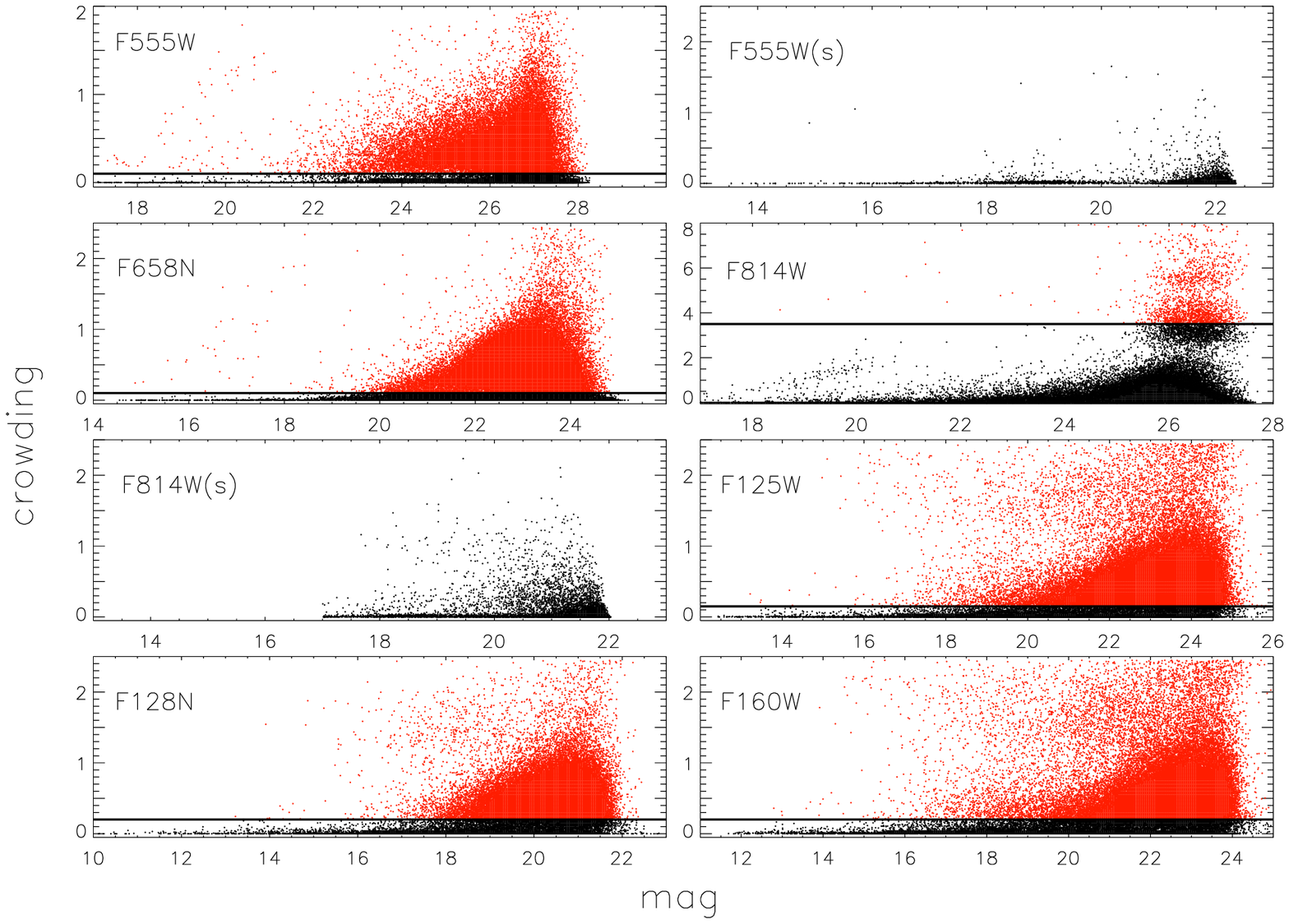}}
\caption{The crowding as a function of magnitude for each filter. The black dots represent the data retained.}
\label{fig:crowding}
\end{figure*}

The crowding is a measure of the flux contamination by surrounding objects. For an isolated star the value is zero. Crowding effects become especially important in very densely populated regions or in areas around very bright saturated stars since the crowding parameter gives a value in magnitudes by how much brighter a star would have been measured if nearby stars would not have been fit simultaneously. Our values chosen to select only those stars with a well-defined PSF and very low contamination can be seen in column 4 of Tab. \ref{tab:cleaning_thresholds}. This selection is visualized in Fig. \ref{fig:crowding}.

\begin{deluxetable}{lrrr}
\tablecaption{The thresholds for cleaning the catalogs \label{tab:cleaning_thresholds}}
\tablewidth{0pt}
\tablehead{
\multicolumn{1}{c}{\textsc{Filter}} & \multicolumn{1}{c}{$\sigma$ [mag]} & \multicolumn{1}{c}{\textsc{Sharpness}} & \multicolumn{1}{c}{\textsc{Crowding}}
}
\tablecolumns{4}
\startdata
$F555W$ 	&$0.5$ & $\pm 0.2$  & $ 0.10 $ \\
$F555W(s)$ 	&$0.5$ & $\pm 0.2$  & $ 2.50 $ \\
$F658N$ 	&$0.5$ & $\pm 0.4$  & $ 0.10 $ \\
$F814W$ 	&$0.5$ & $\pm 0.4$  & $ 3.50 $ \\
$F814W(s)$ 	&$0.5$ & $\pm 0.2$  & $ 2.50 $ \\
$F125W$ 	&$0.2$ & $\pm 0.5$  & $ 0.15 $ \\
$F128N$ 	&$0.2$ & $\pm 0.5$  & $ 0.20 $ \\
$F160W$ 	&$0.1$ & ---		& $ 0.20 $ \\
\enddata
\tablecomments{In this table we give the different thresholds for cleaning the photometric catalogs from spurious objects. The columns 2--4 refer to the quality parameters for point-source photometry as provided DOLPHOT. Objects with values equal to or smaller than the listed thresholds were retained.}
\end{deluxetable}

\subsection{Merging the different filters}
\label{subsec:filter_merging}
For the creation of the final catalog we merged the different filter catalogs. We used the pixel coordinates of the resampled images and we established a maximum radial distance of 1.5~pixels (0.0735 arcsec/pixel) for merging the different filters. This means that the position of an object in two different filters is not allowed to radially deviate by more than 1.5~pixels from each other.

First, we combined the long and the short exposures for the $F555W$ and the $F814W$ band. Then, we used the $F814W$ catalog as basis, and subsequently added one filter after another. At the end, we adopted the World Coordinate System (WCS), in J2000 coordinates, which is provided by the processed long exposures in the $F814W$ filter as the common coordinate system for all exposures.

In order to merge the long and short exposure catalogs this procedure was repeated with a maximum radial distance of 1~pixel, between the same object in the short and long exposures. Objects from the short exposures were only taken into account for magnitudes brighter than 20~mag. For fainter objects, the long exposures were not saturated and offered a higher signal-to-noise (SNR) than the short exposures.

For all objects available in both exposures ($n_{F555W}$=246 and $n_{F814W}$=673, where $n$ is the number of objects) we chose to adopt the point-source photometry with the smaller photometric error which always came, as expected, from the long exposures.

After all filters had been merged into one catalog, all objects that were not detected in at least two filters were deleted. It is unlikely that a star is only detected in one filter. This is not true for the $F160W$ band, however, since it can happen that a very young object emits only far to near-infrared radiation \citep{Lada1984SED} and therefore is only detected in the $H$-band. In order not to lose those (potentially interesting) objects we decided not to remove the $F160W$-only detections.

At the end we have a catalog consisting of objects detected at least in two filters (except for the $F160W$ filter) with a common coordinate system in pixel coordinates, as well as a WCS based on the $F814W$ images.

\subsection{Properties of the photometric catalog}
\label{subsec:cat_properties}
\begin{figure*}[htb]
\resizebox{\hsize}{!}{\includegraphics{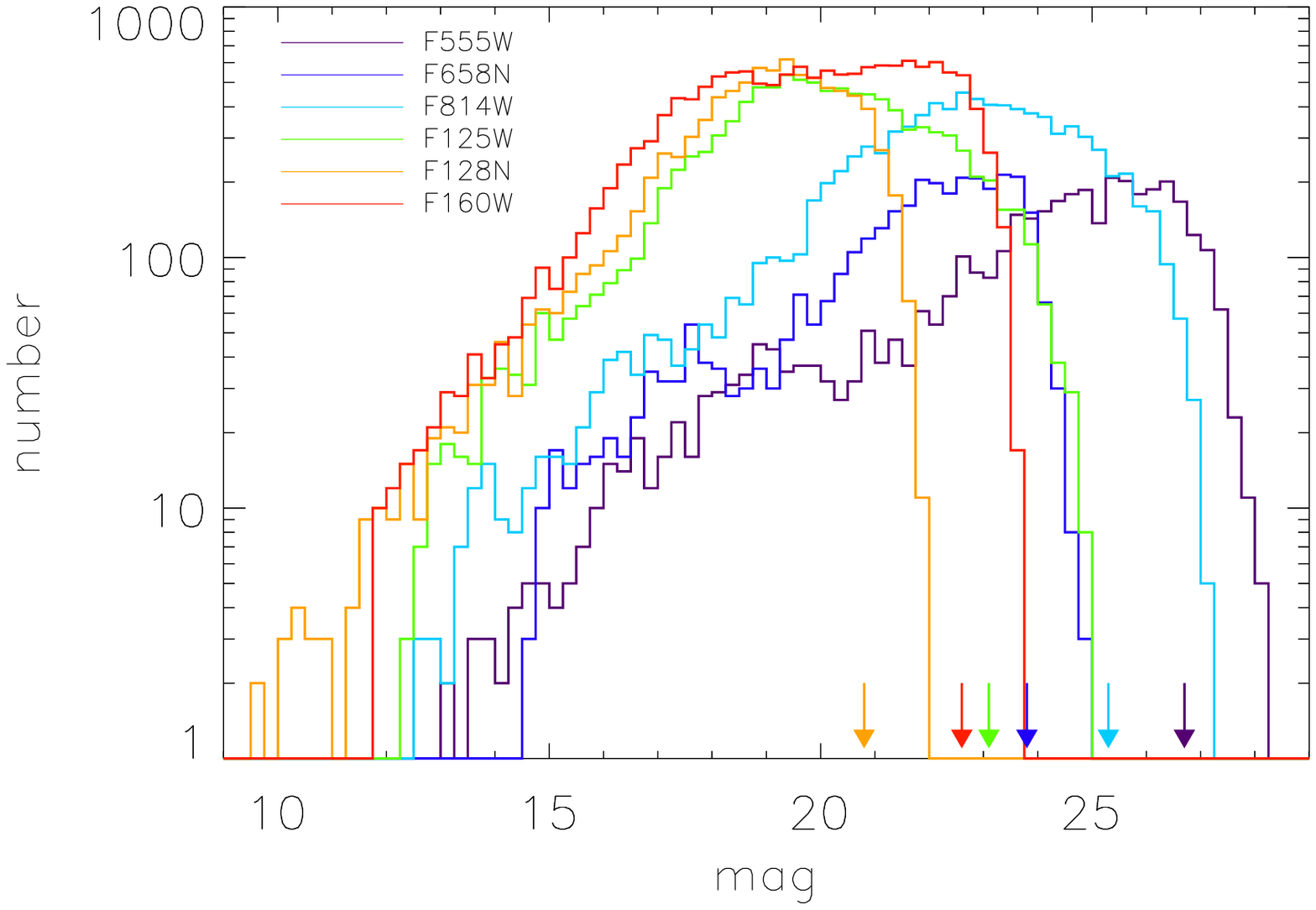}}
\caption{The magnitude distribution of the detections contained in the final Wd2 catalog. The plateau at the faint-end detections in the $F160W$ filter is due to the inclusion of all single-band detections. The arrows mark the limiting magnitudes up to where 90\% of all sources are located (see Tab. \ref{tab:final_catalog}).}
\label{fig:completeness}
\end{figure*}

The final catalog consists of 17\,121 objects in total. For an overview of the magnitude range as well as of the depth of each band, we refer to Fig.\ref{fig:completeness}. The arrows mark the limiting magnitudes up to where 90\% of all sources are located. A summary of the number of point-source detections in the combined catalog is given in Tab. \ref{tab:final_catalog}.

\begin{figure}[htb]
\resizebox{\hsize}{!}{\includegraphics{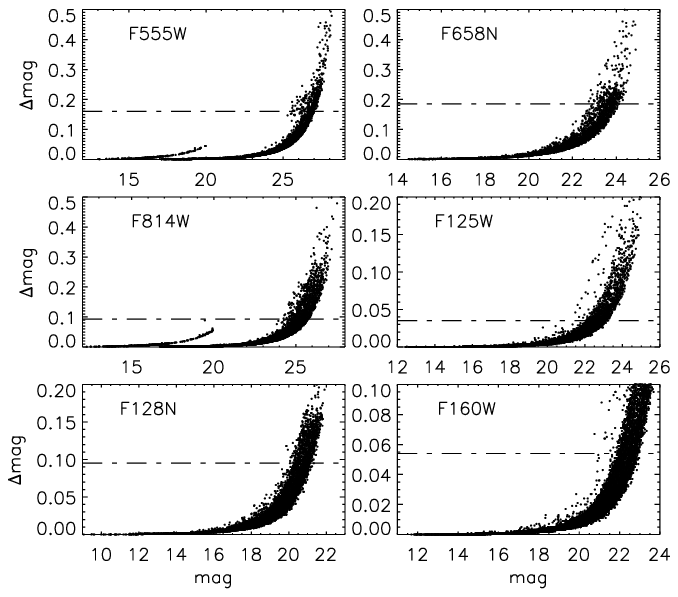}}
\caption{The photometric errors in each filter. The horizontal lines show the limits below which 90\% of all sources are located in the deep exposures (see Tab. \ref{tab:final_catalog}).}
\label{fig:phot_errors}
\end{figure}

The photometric errors are, of course, below the cuts set in the cleaning process. As one can see in Fig. \ref{fig:phot_errors} most sources have much smaller errors than these limits. The limiting errors and resulting magnitudes comprise 90\% of all detected point sources in a given filter when no cuts are imposed, and are given in Tab. \ref{tab:final_catalog}.

\begin{deluxetable}{lrrr}
\tablecaption{Properties of the final catalog \label{tab:final_catalog}}
\tablewidth{0pt}
\tablehead{
\multicolumn{1}{c}{\textsc{Filter}} & \multicolumn{1}{c}{\textsc{Total Number}} & \multicolumn{1}{c}{mag-\textsc{limit} [mag]} & \multicolumn{1}{c}{$\sigma$-limit [mag]}
}
\tablecolumns{4}
\startdata
$F555W$ 	&$3\,922$ & $26.7$  & $0.160$ \\
$F658N$ 	&$3\,556$ & $23.8$  & $0.185$ \\
$F814W$ 	&$9\,281$ & $25.3$  & $0.093$ \\
$F125W$ 	&$10\,981$& $23.1$  & $0.035$ \\
$F128N$ 	&$8\,955$ & $20.8$  & $0.095$ \\
$F160W$ 	&$15\,355$& $22.6$	& $0.054$ \\
\enddata
\tablecomments{In this table we give the main properties of the final photometric catalog. In column 2 we show the total number of all objects per filter. Column 3 and 4 give the limits in magnitudes and the photometric error below which 90\% of all sources are located. In total, 17\,121 objects were detected. 2\,236 point sources had a detection found in all filters.}
\end{deluxetable}

\subsection{The photometric improvement}
\label{subsec:phot_improvement}

The photometric data of our survey are the deepest data obtained for Wd2 so far. The Wide-Field Planetary Camera 2 (WFPC2) \citep{WFPC2} HST data of \citet{Alvarez_13} reach a photometric uncertainty level of 0.2~mag at a depth of $\sim 21.5$~mag and $\sim 21.0$~mag in the $F555W$ and $F814W$ filters, respectively. At the same level of photometric uncertainty we go down to $\sim 27$~mag in the $F555W$ filter and to $\sim 26$~mag in the $F814W$ filter, implying that our data are $>5$~mag deeper in the optical. The $J$ and $H$ photometry by \citet{Ascenso_07} has a limiting magnitude of $\sim 20$~mag and $\sim 18.5$~mag  at a photometric uncertainty limit of $0.1$~mag, respectively, where our HST photometry in the $F125W$ and $F160W$ filter goes down to $\sim 23.5$~mag and $\sim 23$~mag at the same uncertainty levels. Also in the near infrared (NIR) filters our data go $\sim 3$--5~mag deeper than the photometric catalog of \citet{Alvarez_13}. Here it is worth reminding the reader that even when including the 3~s short exposures of the ACS broad-band filters we miss 16 bright objects (see Subsect. \ref{subsec:CMD_morphology}). Those objects are too bright or blended to be accurately measured by DOLPHOT.

The spatial resolution of our dataset is also unprecedented. Especially due to the sub pixel drizzling we are able to reach a resolution of 0.049~arcsec pixel$^{-1}$ in the optical and 0.098~arcsec pixel$^{-1}$ in the IR. This is almost 50\% higher than the pixel resolution of the data of \citet{Ascenso_07} (0.144--0.288~arcsec pixel$^{-1}$) and more than eight times higher than the optical resolution of \citet{Carraro_13} (0.435~arcsec pixel$^{-1}$).

These deep high resolution data in 6 filters give us the unique opportunity to study the properties of Wd2 at an unprecedented level of accuracy and depth. They reveal the young low-mass population (see Sect. \ref{sec:physical_parameters}), give us new information on the spatial distribution of the cluster members (see Sect. \ref{sec:two_clumps}), make it possible to create for the first time a high resolution pixel-to-pixel map of the color excess in order to correct for differential reddening (see Sect. \ref{sec:reddening_map}) and to improve the cluster age determination. 

\section{A high-resolution two-dimensional map of the color-excess}
\label{sec:reddening_map}

Wd2 is located in the Carina-Sagittarius arm of the MW, and thus the local extinction due to interstellar dust is expected to be high. A high resolution, spatially resolved characterization of the extinction in the region is key to a precise determination of age and distance for the Wd2 cluster, and to investigate the existence of an age spread.
Following the methodology outlined by \citet{Pang_11} for NGC3603, we used a combination of H$\alpha$ and Pa$\beta$ images ($F658N$ and $F128N$) to derive a high-resolution, pixel-to-pixel map of the color excess of the gas (g) $E(B-V)_g$ of the Wd2 region.

\subsection{H$\alpha$ and Pa$\beta$ emission}
\label{subsec:emission}

In order to construct the pixel-to-pixel color excess map, we exploited the fact that radiation emitted at shorter wavelengths is absorbed more easily by dust than radiation emitted at longer wavelengths. Therefore, the interstellar dust attenuates the light from a given source emitted in the H$\alpha$ line (rest-frame wavelength $6\,563$~\AA) more than the light emitted in the Pa$\beta$ line (rest-frame wavelength $12\,802$~\AA). Consequently, the observed flux ratio of H$\alpha/$Pa$\beta$ is always smaller than the theoretical (reddening free) ratio computed for the same conditions of electron density and electron temperature.

\citet{Calzetti_96} provides a relation between the color-excess $E(B-V)_g$ of the interstellar gas and the observed ($R_{obs}$) and theoretical ($R_{int}$) flux ratio H$\alpha/$Pa$\beta$ with an assumed extinction law, represented by the total to selective extinction $\kappa(\lambda)=A(\lambda)/E(B-V)$ for a given wavelength $\lambda$:

\begin{equation}
\label{eq:color_excess}
E(B-V)=\frac{-\log{(R_{obs}/R_{int})}}{0.4\left[ \kappa(\mathrm{H}\alpha)-\kappa(\mathrm{Pa}\beta)\right]}.
\end{equation}

\begin{figure*}[htb]
\resizebox{\hsize}{!}{\includegraphics{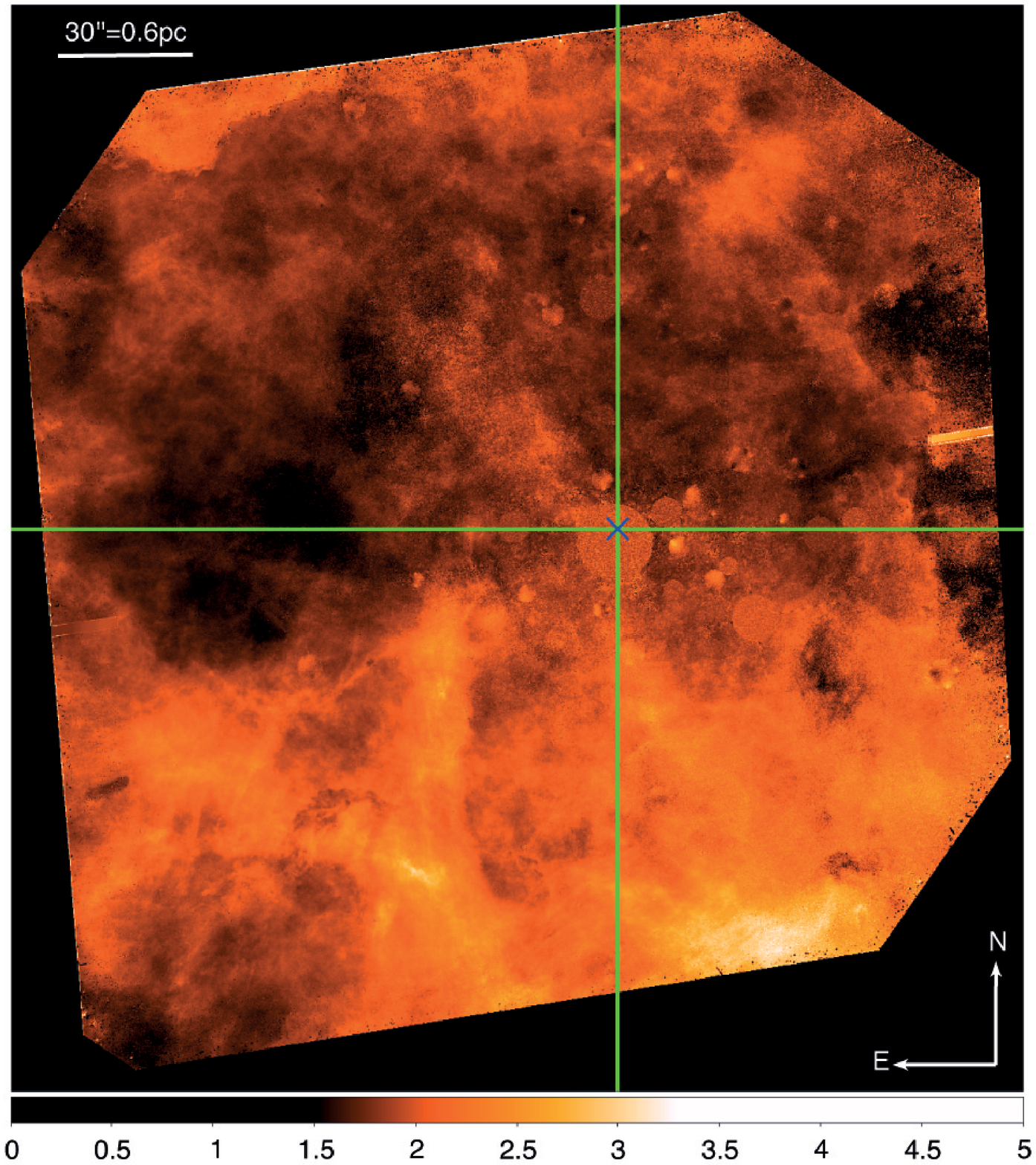}}
\caption{The pixel-to-pixel map of the color excess of the gas $E(B-V)_g$, covering the area where all six filters are available (see Fig. \ref{fig:survey_area}). The cross marks the center of Wd2, as given in Simbad. The green lines indicate the separation in four quadrants (see Sect. \ref{subsec:features_red_map}). $E(B-V)_{g,\rm{min}}=0.002$~mag and $E(B-V)_{g,\rm{max}}=11.7$~mag. The colorbar shows the color excess in magnitudes. All pixels with $E(B-V)_g > 5$ are shown in white. The circular areas near and in the cluster center are caused by the substitution of the saturated pixels by the median of the surrounding area. The two slightly inclined short stripes (one on the left, one on the right side) are caused by the gap between the two ACS chips (see Fig. \ref{fig:survey_area}). They remain at locations in the combined image where only one dither position is available.}
\label{fig:red_map}
\end{figure*}

In order to use Equation \ref{eq:color_excess} we need to get the pixel-to-pixel H$\alpha/$Pa$\beta$ flux ratio from our observations. We applied the following steps in order to properly remove the stellar emission:
\begin{itemize}

\item First, we repeated the drizzling process (see Sec. \ref{sec:observations}) for the ACS images adopting the same pixel size as for the WFC3/IR images ($0.098$~arcsec pixel$^{-1}$). A visual inspection ensured that the images where properly aligned.
\item Due to the use of different instruments covering the whole wavelength range we need all of our images to have the same PSFs. Therefore, we used 40 stars that were non-saturated and without a very close neighbor in each of the six filters in order to determine the PSF. The PSF was largest in the $F128N$ image with a FWHM of $2.26\pm0.312$~pixels. Hence we degraded all the other images to the same resolution by convolving them with a Gaussian whose dispersion ($\sigma^2$) is the difference between the $F128N$ filter ($\sigma^2_{F128N}$) and the one of the filter in question.
\item We calibrated all images in flux by multiplying them by their specific PHOTFLAM value given in the respective header.
\item The continuum flux centered at the H$\alpha$ and Pa$\beta$ wavelength was derived by linear extrapolation of the fluxes in the $F555W$ and $F814W$ ($F125W$ and $F160W$) image pixel-by-pixel at the $F658N$ ($F128N$) wavelength. With such an extrapolation we take into account the slope of the continuum emission of the stars in order to better remove them from the final, pure line-emission images.
\item The interpolated continuum image was then subtracted for properly removing the stars from the $F658N$ ($F128N$) image. Afterwards each pixel value was multiplied by the respective bandwidth resulting in the pure line emission image as well as the $R_{obs}$ flux ratio.
\item For the theoretical flux ratio of H$\alpha$/Pa$\beta$, $R_{int}$, we derived a value of $17.546$ with an assumed electron temperature of 10\,000~K and an electron density of 100~cm$^{-3}$ \citep{Osterbrock_89}. We adopted a normal total-to-selective extinction of $R_V=3.1$. From the extinction law of \cite{Cardelli_89} we derived $\kappa(\mathrm{H}\alpha)=2.437$ and $\kappa(\mathrm{Pa}\beta)=0.8035$. With \citet{Fitzpatrick_99} we derived $\kappa(\mathrm{H}\alpha)=2.346$ and $\kappa(\mathrm{Pa}\beta)=0.7621$.
\item Finally we were able to apply Equation \ref{eq:color_excess} to construct the pixel-to-pixel map of $E(B-V)_g$ shown in Figure \ref{fig:red_map}.

\end{itemize}

\subsection{The features of the color excess map}
\label{subsec:features_red_map}

The two-dimensional color-excess map (see Fig. \ref{fig:red_map}) is created with a spatial resolution ($0.098$~arcsec pixel$^{-1}$) that has never been reached before for this region. The very inhomogeneous distribution of the dust extinction shows that differential reddening must be taken into account before any conclusion can be reached on the physical parameters of Wd2. We will use this map to get a dereddened cluster population, where for each of the stars the individual reddening is estimated locally (see Sect \ref{sec:physical_parameters}).

\begin{figure*}[htb]
\resizebox{\hsize}{!}{\includegraphics{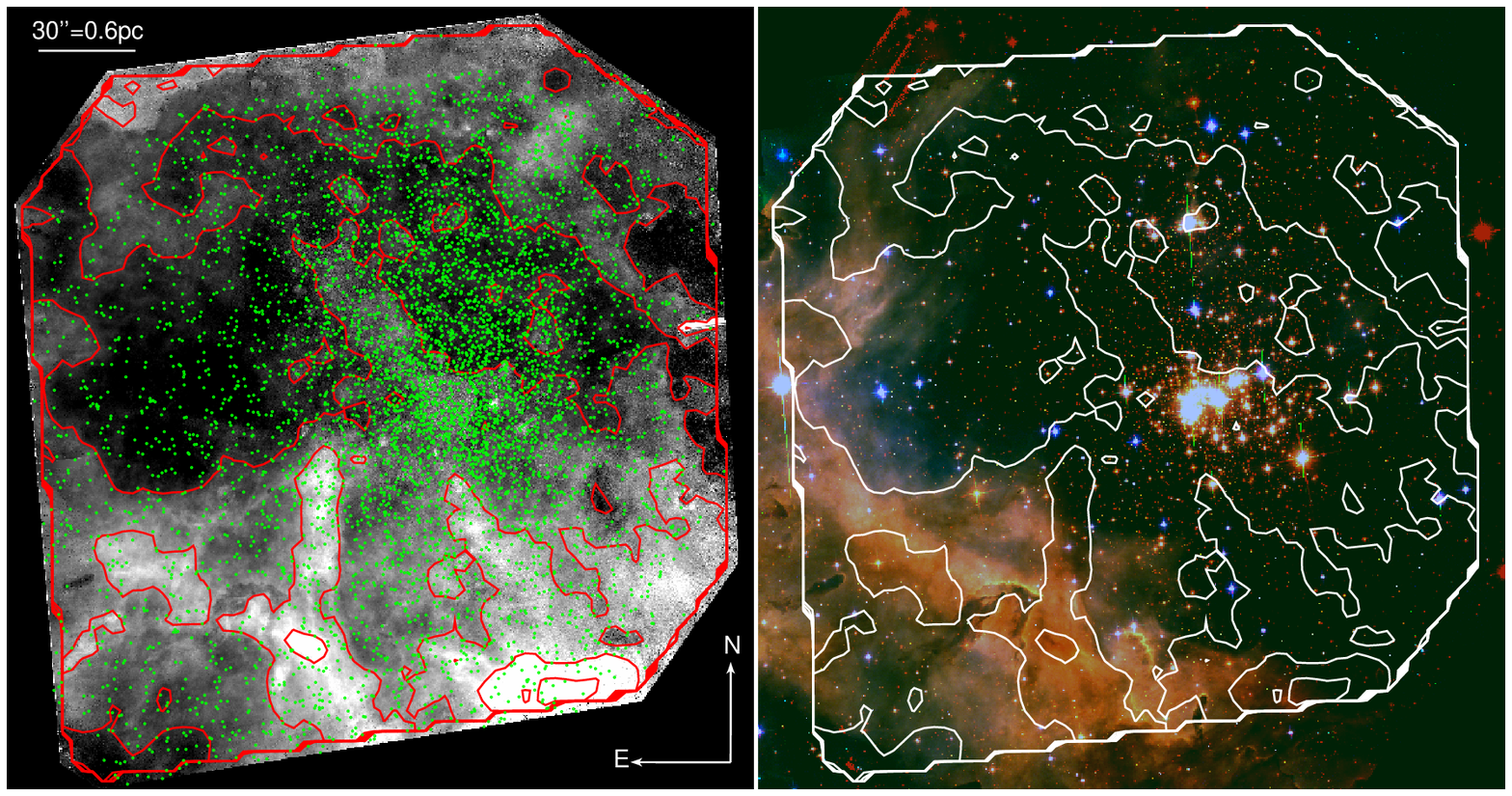}}
\caption{\textbf{Left: }The pixel-to-pixel map of the color excess $E(B-V)_g$ covering the area where all six filters are available (see Fig. \ref{fig:survey_area}). The green points are the stars in RCW~49 (see Sect. \ref{sec:physical_parameters}), and the contours represent $E(B-V)_g$ in the range of 0 to 2.86 in 9 linear steps. \textbf{Right: }RGB composite image of the $F125W$ (red), $F814W$ (green), and $F555W$ (blue) filter.}
\label{fig:red_map_RGB}
\end{figure*}

In Fig. \ref{fig:red_map_RGB}, the reddening map is shown on the left in black and white over-plotted with the selected cluster members (see Sect. \ref{sec:physical_parameters}) in green, and the contours representing $E(B-V)_g$. On the right we show the RGB image of the same spatial extent over-plotted with the same $E(B-V)_g$ contours. This image emphasizes the location of the stellar population of Wd2 with respect to the highly variable color excess on small scales. Here it becomes clear that this map will improve the correction for differential reddening of each star individually.

The color excess map is morphologically very inhomogeneous. In the center where the cluster is located (marked by a cross in Fig. \ref{fig:red_map}), $E(B-V)_g$ has a value of $\sim 1.8$~mag. Here we also see most of the very bright stars in the color composite image (see Fig. \ref{fig:red_map_RGB}). Also in the very central region around the cluster there still seems to be a concentration of gas and dust, although we cannot say whether these components are mainly located in front of the cluster as seen along our line of sight, or in the cluster. A "bridge" with a slightly increased $E(B-V)_g \approx 1.9$~mag extends from the cluster center to the North-East. In Fig. \ref{fig:red_map_RGB} one can see that there are no massive luminous stars visible within this "bridge". This "bridge" is not visible in the RGB image. Towards the South and the East where a higher concentration of dust and gas is observed, the color excess increases to values larger than 2.0~mag. In very dense regions it even can exceed 2.5~mag. The color excess map follows very well the morphology of the gas clouds showing a direct connection between the color excess $E(B-V)_g$ and the cloud structure. As expected the color excess is higher where the clouds are denser.

\begin{figure}[htb]
\resizebox{\hsize}{!}{\includegraphics{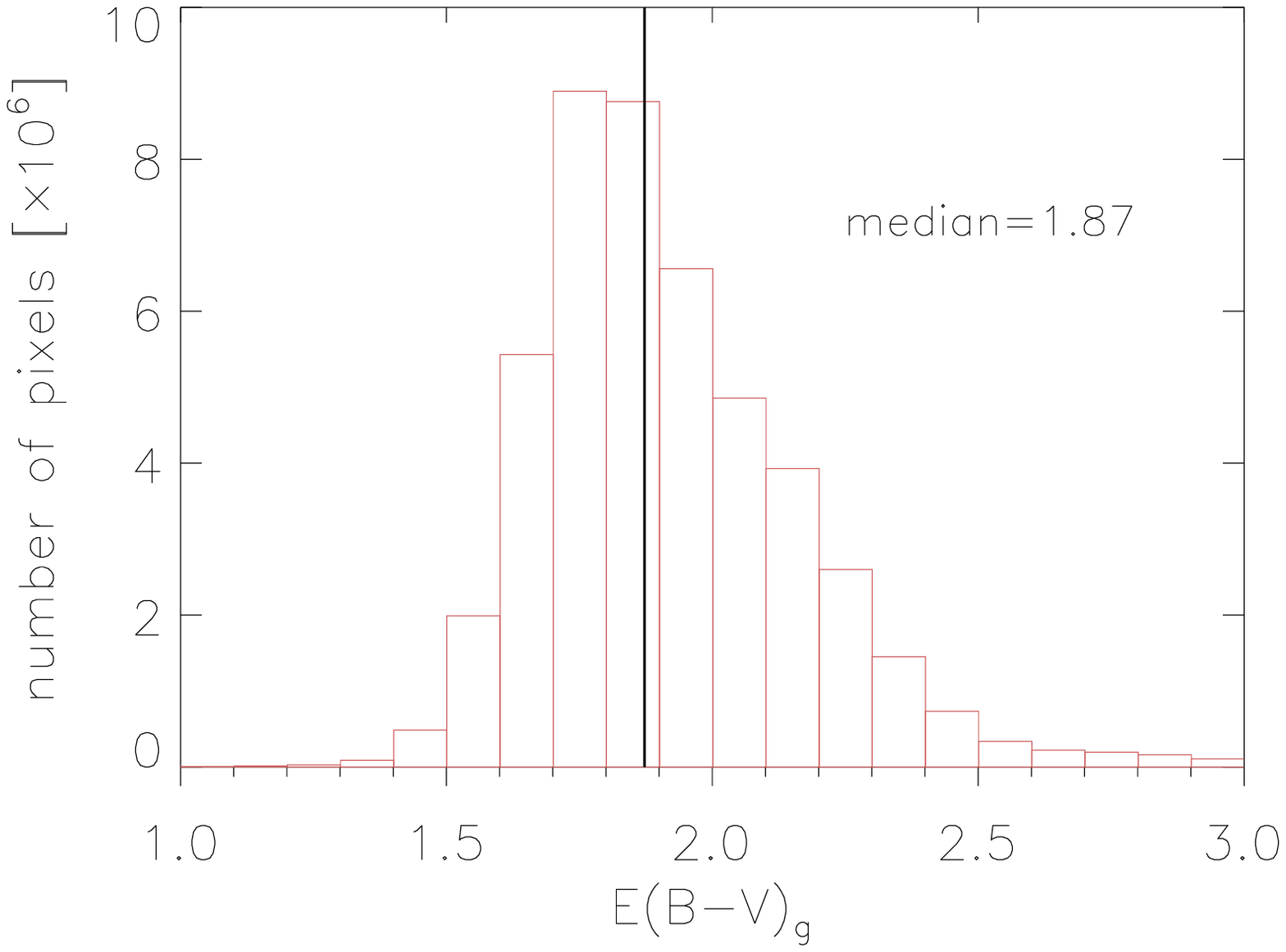}}
\caption{The distribution of the color excess $E(B-V)_g$ with a median of 1.87~mag. Only 10\,742 pixels (0.23\%) have a value of $E(B-V)_g>3.0$~mag and only 1\,018 pixels (0.02\%) exceed 5.0~mag}
\label{fig:excess_distribution}
\end{figure}

The median reddening of the pixels across the reddening map is $E(B-V)_g=1.87$~mag. The vast majority of all pixels have $E(B-V)_g$ between 1--3~mag. Only 10\,742 pixels (0.23\%) have $E(B-V)_g>3.0$~mag and only 1\,018 pixels (0.02\%) exceed 5.0~mag. The distribution of the nebular color excess values can be seen in Fig. \ref{fig:excess_distribution}.

\begin{figure}[htb]
\resizebox{\hsize}{!}{\includegraphics{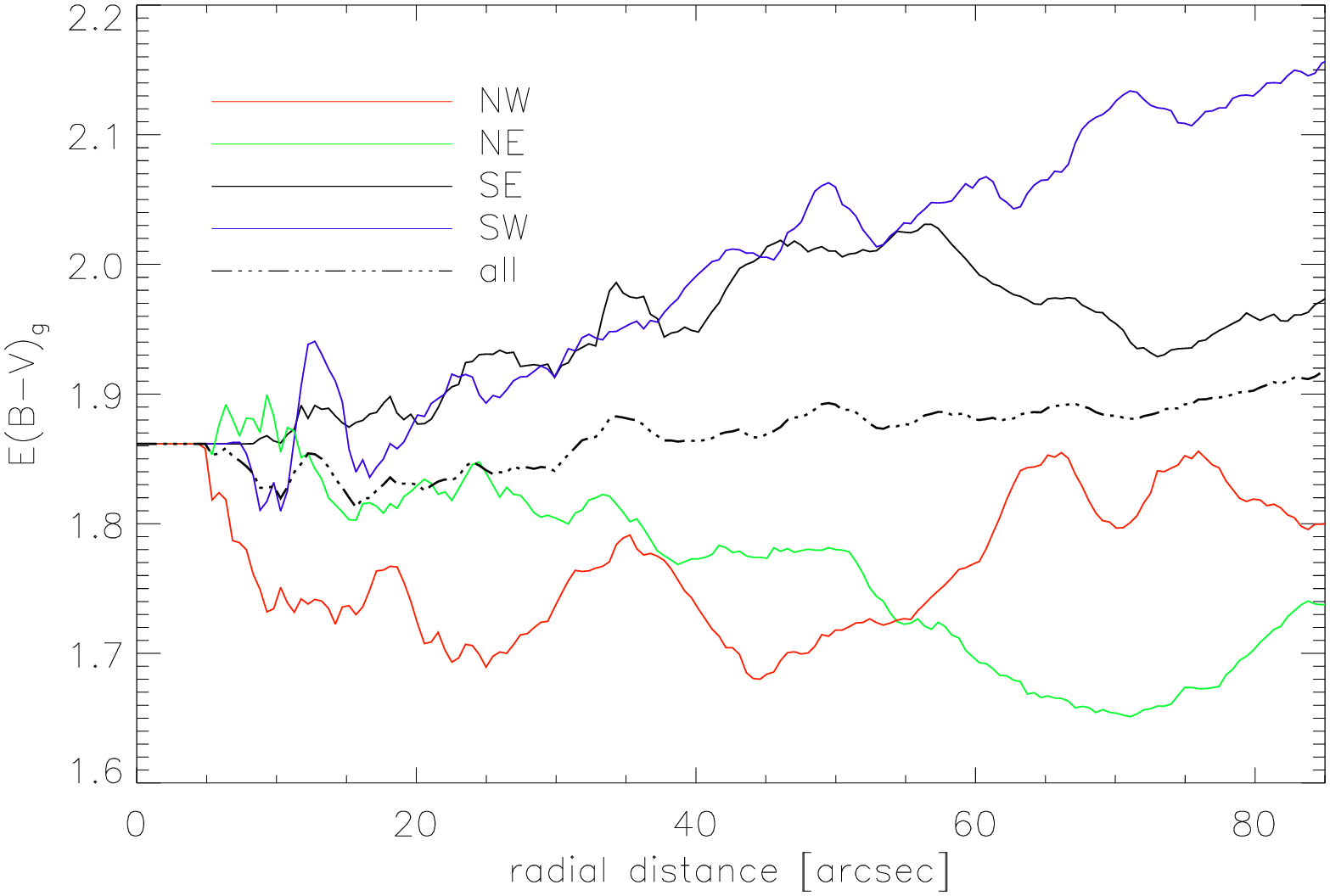}}
\caption{The averaged radial distribution of the color excess $E(B-V)_g$ in the four quadrants around the cluster center (cross in Fig. \ref{fig:red_map}) including the average over all directions (dashed-dotted line).}
\label{fig:differential_reddening}
\end{figure}

As shown in Fig. \ref{fig:red_map} the range of $E(B-V)_g$ is quite large. We carried out an analysis of the radial distribution of the color excess. We divided the color excess map into four quadrants (NW, NE, SW, SE) (see Fig. \ref{fig:red_map}), with the origin at the center of the cluster (cross in Fig. \ref{fig:red_map}). In Fig. \ref{fig:differential_reddening} one can see the azimuthal dependence of the color excess. The dashed-dotted line shows the average of all quadrants. To the South the color excess increases almost constantly from $\sim 1.86$~mag to 2.15~mag in the South-West and $\sim 2.0$~mag in the South-East. To the North, $E(B-V)_g$ decreases to $\sim 1.7$~mag.

\subsection{Uncertainties of the color excess map}
\label{subsec:errors_red_map}
Before applying the reddening map to our photometric measurements, we needed to assess the accuracy of the map itself. We did that by assuming that the electron counts on the detector follow a Poisson distribution. We started with a Poisson error of the counts per pixel and performed a proper error propagation throughout our whole computation, beginning with the continuum interpolation up to the final $E(B-V)_g$ from Equation \ref{eq:color_excess}.

\begin{figure}[htb]
\resizebox{\hsize}{!}{\includegraphics{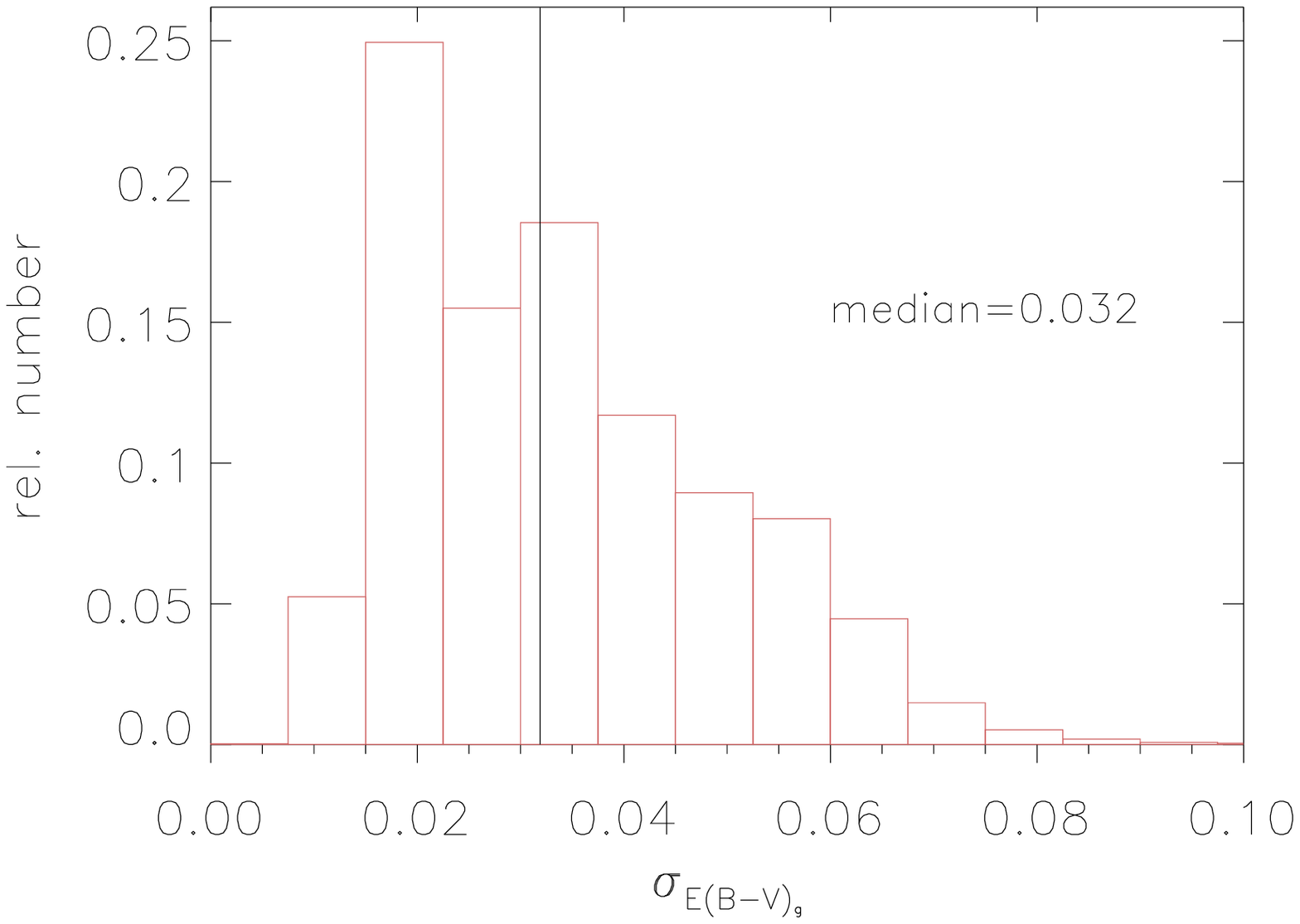}}
\caption{The distribution of $\sigma_{E(B-V)_g}$ with a binsize of 0.0075~mag. The median value is 0.032~mag. The number is normalized to the total number of pixels.}
\label{fig:sigma_color_excess}
\end{figure}

In Fig. \ref{fig:sigma_color_excess} we show the distribution of $\sigma_{E(B-V)_g}$ with a median of 0.032~mag. 90\% of all pixels have an error smaller than 0.057~mag in the inferred nebular color excess.

As mentioned before, the reddening map depends on the adopted extinction law. For all further analyses we used the extinction law of \citet[][hereafter denoted with the subscript C]{Cardelli_89}. For the sake of comparison, we also created the color excess map with the extinction law of \citet[][hereafter denoted with the subscript FP]{Fitzpatrick_99}. Both extinction laws are almost identical in the optical regime and so are the results. The median of the difference between $E(B-V)_{FP}$ and $E(B-V)_{C}$ is $0.05\pm0.018$~mag.

A non-negligible effect is the contamination of the wide-band filters with emission lines (e.g.: \ion{O}{3}, \ion{S}{3}, H$\beta$ ). Due to a lack of spectra of the gas and the large differences in published line emissions between different \ion{H}{2} regions, we are unable to estimate the amount of contamination due to those lines. Compared to the Pa$\beta$ line emission we expect this contamination in the wide-band filters to be small.

\begin{figure}[htb]
\resizebox{\hsize}{!}{\includegraphics{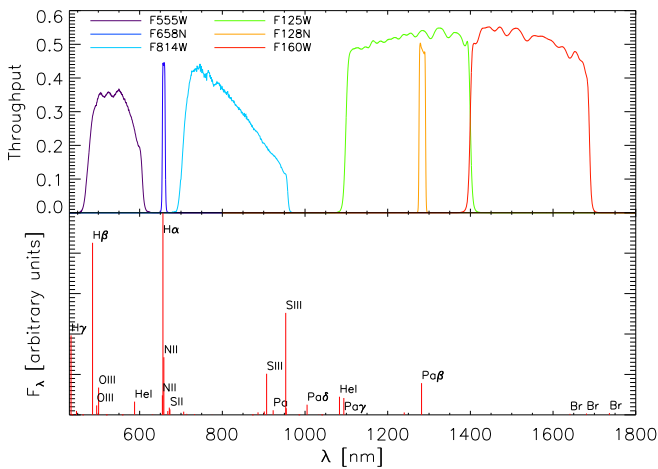}}
\caption{\textbf{Top:} Shown are the throughputs of the six used filters, derived with Synphot. The $F128N$ passband is located within the wavelength range covered by the $F125W$ filter. Therefore, the Pa$\beta$ emission line contributes to the $F125W$ continuum. \textbf{Bottom:} Synthetic spectral lines from the \ion{H}{2} Regions Library \citep{Panuzzo_03} show the emission features contaminating the continuum filters. The H$\alpha$ line is truncated for plotting reasons.}
\label{fig:throughputs}
\end{figure}

In order to visualize the possible contamination of the color excess map by nebular line emission we show in Fig. \ref{fig:throughputs} the throughputs (top panel) of our filters together with the simulated spectral lines (bottom panel) of a typical \ion{H}{2} region using the \ion{H}{2} Regions Library\footnote{\url{http://pasquale.panuzzo.free.fr/hii/}} based on the code CLOUDY \citep{Panuzzo_03}. For the model parameters we adopted the values ($n_{\rm{H}} \sim 100$~cm$^{-3}$, $Q_0=4.2 \cdot 10^{51}$~s$^{-1}$) of \citet[][and references therein]{Pellegrini_11} for the massive star cluster \object{R136}, since its content of massive stars is comparable to Wd2. Since we are only qualitatively interested in the possible gas emission lines the adopted values fulfill our needs.

The Pa$\beta$ line is the only line for which we have information in our survey. H$\alpha$ is located outside the broadband filters and, therefore, not contaminating the continuum. Assuming the flux density to be the same in $F125W$ and $F128N$, we can derive the contribution of Pa$\beta$ via the flux ratio $F125W/F128N$. Both filters were multiplied by their respective bandwidth and PHOTFLAM value. We estimate the contamination of the continuum by the Pa$\beta$ to be $\sim 22\%$.

Taking into account results from \citet{Pang_11} we estimate the contamination of the other possible emission lines in the other filters to be about 10-15\% which could increase the overall color excess by $\sim$0.1-0.15~mag.

Finally, the $E(B-V)_g$ excess is also affected by the electron density and temperature used to calculate the $R_{int}$ ratio. We calculated $E(B-V)_g$ for different electron densities and temperatures. For an increase of the electron density by a factor of two, $E(B-V)_g$ decreases by 0.00124~mag. For the extreme cases of $N_e=10\,000 (100)$~cm$^{-3}$ and $T_e=5\,000 (20\,000)$~K, $E(B-V)_g$ changes by $+0.028(-0.034)$~mag.

In summary, we can say that the uncertainties in electron density and temperature are introducing a small spread, which is below the 90\% limit of the photometric uncertainty of the different filters (see Fig. \ref{fig:phot_errors}). The unknown line contamination, on the other hand, introduces a possible increase of the color excess of the ga,s $E(B-V)_g$ of $\sim$0.1-0.15~mag. This shift can be calibrated out when transforming $E(B-V)_g$ to the stellar color excess $E(B-V)_\star$ (see eq. \ref{eq:gas_excess_transformation}).

\subsection{Limitations of the color excess map}
\label{subsec:properties_red_map}

The removal of the stellar contamination by interpolating and subtracting the continuum emission worked well for all objects up to a certain brightness limit. Problems occurred for saturated objects and their spikes as well as for those objects with diffraction rings. At the positions of these sources, the pixel values are not reliable anymore. Therefore, the interpolation and thus the subtraction might partly or totally fail for these objects.

For small areas with just a few unreliable pixel values we were able to clean the final reddening map by performing a spline interpolation. This effect appears for objects close to the saturation limit meaning that only a few pixels are saturated or the object is saturated just in one filter.

For cosmic rays, dead and hot pixels, being these cases where just a single pixel is defective, we were able to clean the final map by replacing the affected pixels by the median value of their surrounding.

For the very bright objects as well as for the whole cluster center, the area of saturated pixels was too large for a meaningful interpolation. Here we replaced the entire affected area by the median reddening calculated over an annulus around that area. For the map in Fig. \ref{fig:red_map} we additionally introduced a random Poisson noise to the interpolated pixel values (see Fig. \ref{fig:phot_errors}). Those values should thus be handled with caution.

\section{The stellar reddening towards the direction of Wd2}
\label{sec:stellar_reddening}
In order to get a better insight into the physical parameters of Wd2 we need to derive the variability of the reddening for the stellar population of the cluster.

\subsection{Transformation of the gas excess map to the stellar excess map}
\label{subsec:excess_transformation}

After creating the pixel-to-pixel gas excess map $E(B-V)_g$ there was the need to translate this color excess into a stellar color excess $E(B-V)_\star$ in order to properly deredden the stellar photometry. \citet{Calzetti_97b} and \citet{Calzetti_00} found a linear empirical relation between the stellar and the gas reddening ($E(B-V)_\star \sim f\cdot E(B-V)_g$). In star forming regions, the dust in front of the stars is removed or destroyed by stellar feedback. Therefore, the stellar reddening is smaller than the reddening of the gas. The interstellar medium (ISM) consists of a different composition (grain size, density, etc.) throughout the Galaxy. Therefore, this linear, empirical relation is only valid for a small range in color excess $E(B-V)$ and needs to be determined for each cluster independently.

We used the results of the ground based optical spectral types of massive stars in Wd2 from \citet{Alvarez_13} and \citet{Rauw_07,Rauw_11} to obtain the bolometric magnitude of each of those stars using the Tab~1.4 of \citet{Sparke_07}. Combining this information with the zero-age-main-sequence (ZAMS)\footnote{The ZAMS was derived, combining PARSEC isochrones with the information given on the webpage, at which age a star for a certain mass interval reaches the main sequence.} derived from the isochrones of the \textsc{PAdova and TRieste Stellar Evolution Code}\footnote{\url{http://stev.oapd.inaf.it/cmd}} \citep[hereafter: PARSEC,][]{Bressan_12} we could obtain the theoretical intrinsic colors in the proper filter set of the WFPC2. We used this information to determine the color excess of the star combining the intrinsic color and the color derived with the photometric catalog of \citet{Alvarez_13} ($E(B-V)_\star=(B-V)_\star-(B-V)_{th}$) since our photometric catalog does not cover the $B$-band. We then used the proper transformation of the photometry from the WFPC2 photometric system to the Johnson-Cousins system:

\begin{eqnarray}
\label{eq:color_excess_transformation}
\frac{E(\lambda_1-\lambda_2)_g}{E(B-V)_g}= \nonumber \\
\frac{a_{\lambda_1}+\frac{b_{\lambda_1}}{R_V}-a_{\lambda_2}-\frac{b_{\lambda_2}}{R_V}}{a_{B}+\frac{b_{B}}{R_V}-a_{V}-\frac{b_{V}}{R_V}}.
\end{eqnarray}

The parameters $a$ and $b$ are calculated with Eq. 3a and 3b from \citet{Cardelli_89} and given in Tab. \ref{tab:color_excess_transformation}. $\lambda_1$ and $\lambda_2$ are the pivot wavelengths of the $F439W$ and $F555W$ filters, respectively. For the total-to-selective extinction, we used the same value as the color excess map ($R_V=3.1$). We should note here that even though we infer an $R_V$ here in order to determine $E(B-V)_\star$ even that we analyze $R_V$ in Subsect. \ref{subsec:R_V}. A variability of $R_V$ between 3.1 and 4.0 leads to an change in $E(B-V)_\star$ of $\sim 0.6\%$. In Table \ref{tab:comp_spec_data}, we list the values of the gas excess from our gas excess map, as well as the calculated stellar excess, (Column 8-11) for each star.

\begin{deluxetable}{crrrr}
\tablecaption{The transformation parameters \label{tab:color_excess_transformation}}
\tablewidth{0pt}
\tablehead{
\multicolumn{1}{c}{\textsc{Filter}} & \multicolumn{1}{c}{\textsc{Filter system}} &\multicolumn{1}{c}{$\lambda_P$ [nm]} & \multicolumn{1}{c}{a} & \multicolumn{1}{c}{b}
}
\tablecolumns{5}
\startdata
$F336W$ & WFPC2	 & $334.44$ & $0.87338$   &   $2.37152$  \\
$F439W$ & WFPC2	 & $431.13$ & $0.99552$   &   $1.12082$  \\
$B$			 & ---	 & $445$\footnote{The pivot wavelength of the $B$ and $V$ band are taken from \citet[][p.53, Table 2.1]{Binney_98}} 	& $1.00213$    	 &   $0.94137$  \\
$V$	 	 	& ---	 & $551$\footnotemark[1]  	& $1$    		  &   $0$  		 \\
$F555W$ & ACS	 & $536.10$	& $1.00699$	  &   $0.06883$  \\
$F555W$ & WFPC2	 & $543.90$	& $1.00311$	  &   $0.02704$  \\
$F814W$ & WFPC2	 & $801.22$	& $0.78049$	  &   $-0.57273$  \\
$F814W$ & ACS	 & $805.70$	& $0.77652$	  &   $-0.58020$  \\
$F125W$ & WFC3IR & $1248.6$	& $0.40149$	  &   $-0.36821$  \\
$F160W$ & WFC3IR & $1536.9$	& $0.28735$	  &   $-0.26382$  \\

\enddata
\tablecomments{The parameters $a$ and $b$ for transforming the color excess of the stars $E(\lambda_1-\lambda_2)$ to an excess $E(B-V)$. The parameters $a$ and $b$ are calculated with Eq. 3a and 3b from \citet{Cardelli_89}.}
\end{deluxetable}

In order to get the transformation between the stellar excess and the gas excess we used all stars with spectroscopy that have individual reddening values (e.g.: are not located in the center) and are main-sequence stars (class V without special line features).  A linear, error weighted regression revealed the following relation between the gas and the stellar extinction:

\begin{eqnarray}
\label{eq:gas_excess_transformation}
E(B-V)_\star =0.4314 \cdot E(B-V)_g + 0.7400.
\end{eqnarray}

\begin{figure}[htb]
\resizebox{\hsize}{!}{\includegraphics{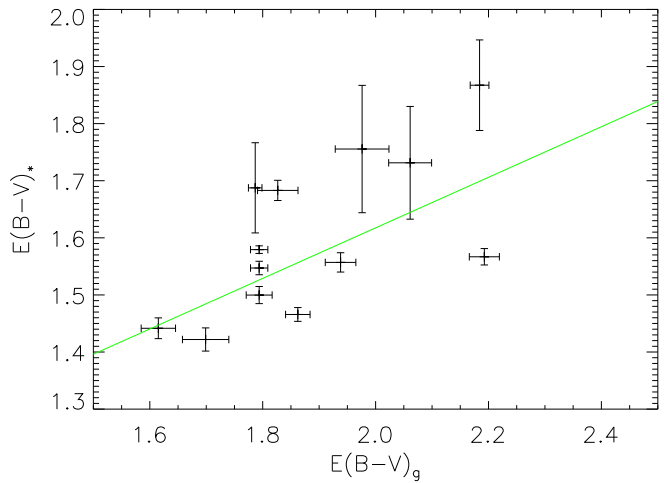}}
\caption{The 13 stars with spectroscopic data used to calculate the transformation between the pixel-to-pixel gas excess map $E(B-V)_g$ to the stellar excess $E(B-V)_\star$ are shown here. The green line is the best fit linear regression including error weighting.}
\label{fig:stellar_gas_excess}
\end{figure}

It is satisfactory to use a total-to-selective extinction of $R_V=3.1$ for creating the color-excess map of the gas (see Sect. \ref{sec:reddening_map}) since a different $R_V$ will be compensated by different regression parameters in Eq. \ref{eq:gas_excess_transformation} and, therefore, leaves the overall outcome for $E(B-V)_\star$ unaffected. This results in an intersect greater than zero. Eq. \ref{eq:gas_excess_transformation} now allows us to translate the $E(B-V)_g$ gas extinction map into a map of the stellar extinction $E(B-V)_\star$. The overall median of $E(B-V)_g=1.87$~mag transforms to $E(B-V)_\star=1.55$~mag. In Fig. \ref{fig:stellar_gas_excess} one can see the stars used to calculate the transformation.

\begin{figure}[htb]
\resizebox{\hsize}{!}{\includegraphics{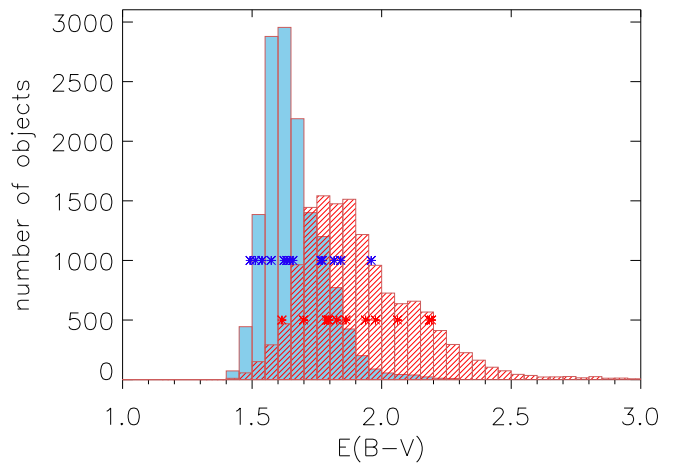}}
\caption{The distribution of the color excess $E(B-V)$ for each of the 14\,199 objects in the catalog, covered by the color excess map. The gas excess distribution is shown in red while the stellar excess distribution is shown in blue. The asterisks mark the color excesses of the thirteen stars with spectral types used to obtain the linear regression for transforming the gas excess into a stellar excess.}
\label{fig:stellar_gas_excess_distribution}
\end{figure}

In Fig.\ref{fig:stellar_gas_excess_distribution} the histogram of the distribution of the color excess $E(B-V)$ is plotted for each of the 14\,199 objects in the catalog spatially covered by the color excess map. The gas excess distribution is shown in red while the stellar excess distribution is shown in blue. The asterisks mark the color excesses of the thirteen stars with spectral types, which we used to obtain the linear regression for transforming the gas excess into a stellar excess. It can be seen that the color excess of the 13 stars is equally distributed over the whole range in color excess of all objects in the catalog. Therefore, the linear approach is valid for the complete catalog. However, we caution that this transformation cannot be used to extrapolate to $E(B-V)_g=0$~mag or to very high $E(B-V)_g$ values.

\begin{deluxetable*}{rrcrrrrrrrrrc}
\tablecaption{Comparison to spectroscopic data \label{tab:comp_spec_data}}
\tablewidth{0pt}
\tabletypesize{\scriptsize}
\tablehead{
\multicolumn{1}{c}{\textsc{RA}} &\multicolumn{1}{c}{\textsc{Dec}}  &\multicolumn{1}{c}{\textsc{Spec. Type}} & \multicolumn{1}{c}{$F555W$} & \multicolumn{2}{c}{$(F439W-F555W)$}  & \multicolumn{3}{c}{$E(B-V)_\star$}  & \multicolumn{2}{c}{$E(B-V)_g$} &  \multicolumn{1}{c}{\textsc{An.}}\\
\multicolumn{1}{c}{\textsc{J2000}} &\multicolumn{1}{c}{\textsc{J2000}}  &\multicolumn{1}{c}{} & \multicolumn{1}{c}{} & \multicolumn{1}{c}{int,th} &  \multicolumn{1}{c}{V.-A.} & \multicolumn{1}{c}{\textsc{TCD}} & \multicolumn{1}{c}{map} & \multicolumn{1}{c}{$\sigma_{\rm{map}}$} & \multicolumn{1}{c}{map} & \multicolumn{1}{c}{$\sigma_{\rm{map}}$} & \multicolumn{1}{c}{}
}
\tablecolumns{13}
\startdata
10:23:55.176 &  -57:45:26.89     & O4 V					 & 15.610		& -0.281  &		    1.776 & 1.841	& 1.526	 &    0.013 &    1.822  &      0.024		&   b\\
10:23:56:160 &  -57:45:29.99     & O4 V--III((f))	   & 14.514		& -0.281  &		    1.451 & ---		& 1.488  &    0.006 &    1.734  &      0.020		&   b\\
10:23:59.201 &  -57:45:40.53   	 & O7.5 V				& 16.015	& -0.270  &		    1.666 & 1.734	& 1.629	 &    0.099 &    2.061  &      0.038		&   b\\
10:24:01.920 &  -57:45:32.62  	 & O8.5 V				& 16.100	& -0.263  &		    1.489 & 1.574	& 1.686	 &    0.014 &    2.193  &      0.027		&  	a,b \\
10:24:00.353 &  -57.45:42.71	 & O8 V		 			& 15.863	& -0.267  &		    1.615 & 1.687	& 1.528	 &    0.018 &    1.827  &      0.036		&   b\\
10:23:00.480 &  -57:45:24.01     & O4 V	 				& 14.546	& -0.281  &		    1.485 & 1.581	& 1.514	 &    0.007 &    1.794  &      0.015		&   b\\
10:24:00.499 &  -57.44:44.53   	 & B1 V					& 15.587	& -0.254  &		    1.358 & 1.441	& 1.437	 &    0.018 &    1.615  &      0.030		&   b\\
10:24:00.713 &  -57.45:25.42     & O8 V					& 15.466	& -0.267  &		    1.410 & 1.503	& 1.514	 &    0.015 &    1.794  &      0.023		&   b\\
10:24:00.816 &  -57.45:25.87     & O6.5 V				& 14.866	& -0.276  &		    1.454 & 1.549	& 1.514	 &    0.012 &    1.794  &      0.015		&   b\\
10:24:00.979 &  -57.45:05.50     & B1 V					& 16.111	& -0.254  &		    1.336 & 1.421	& 1.473	 &    0.020 &    1.699  &      0.041	 	&   a,b\\
10:24:01.070 &  -57.45:45.73     & O9.5 V				& 16.239	& -0.259  &		    1.482 & 1.561	& 1.576	 &    0.017 &    1.938  &      0.027		&   b\\
10:24:01.200 &  -57:45:31.07     & O3 V	 				& 13.487	& -0.283  & 	    1.513 & 1.608	& 1.543  &    0.017 &    1.862  &      0.006		&   \\
10:24:01.392 &  -57:45:29.66     & O4 V	 				& 14.084	& -0.281  &		    1.411 & 1.515	& 1.543	 &    0.006 &    1.862  &      0.007		&   \\
10:24:01.454 &  -57.45:31.33     & O3 V					& 15.042	& -0.283  &		    1.403 & 1.510	& 1.543	 &    0.007 &    1.862  &      0.017		&   \\
10:24:01.524 &  -57.45:57.06  	 & O6 V					& 15.053	& -0.278  &		    1.810 & 1.868	& 1.682	 &    0.079 &    2.184  &      0.017		&  	b\\   
10:24:01.610 &  -57.45:27.89     & O5.5 V				& 14.921	& -0.278  &		    1.378 & 1.482	& 1.543	 &    0.008 &    1.862  &      0.027 		&   \\ 
10:24:01.889 &  -57.45:40.05     & O9.5 V				& 16.574	& -0.259  &		    1.656 & 1.717	& 1.543  &    0.023 &    1.862  &      0.042		& 	a \\
10:24:01.889 &  -57.45:28.00     & O8 V	 				& 15.524	& -0.267  &		    1.510 & 1.593	& 1.543  &    0.015 &    1.862  &      0.022		&    \\
10:24:02.064 &  -57:45:28.01     & O6 III	 			& 14.453	& -0.278  &		    1.664 & ---		& 1.543	 &    0.010 &    1.862  &      0.009		&   \\
10:24:02.186 &  -57.45:31.32  	 & O9.5 V				& 16.609	& -0.259  &		    1.601 & 1.668	& 1.543	 &    0.020 &    1.862  &      0.040		&  	a\\
10:24:02.256 &  -57.45:35.12   	 & O4--5 V			& 14.736	& -0.281  &		    1.498 & 1.592	& 1.543	 &    0.007 &    1.862  &      0.018		&    \\
10:24:02.304 &  -57:45:35.53     & O4.5 V	 			& 13.893	& -0.281  &		    1.475 & 1.571	& 1.543	 &    0.006 &    1.862  &      0.008		&   \\
10:24:02.376 &  -57:45:30.64     & O3 V((f)) 			& 13.878	& -0.283  &		    1.618 & ---		& 1.543	 &    0.018 &    1.862  &      0.004		&   \\
10:24:02.414 &  -57.45:47.11  	 & O9.5 V				& 16.637	& -0.259  &		    1.704 & 1.760	& 1.592	 &    0.112 &    1.976  &      0.048		&  	a,b \\
10:24:02.448 &  -57.44:36.13     & O5 V--III			& 13.155	& -0.280  &		    1.359 & ---		& 1.479	 &    0.052 &    1.712  &      0.003		&  	b \\
10:24:02.518 &  -57.45:31.47     & O8.5 V	 			& 15.790	& -0.263  &		    1.598 & 1.671	& 1.543	 &    0.014 &    1.862  &      0.033		&  	 \\
10:24:02.555 &  -57.45:30.52     & O8.5 V				& 16.243	& -0.263  &		    1.649 & 1.717	& 1.543	 &    0.017 &    1.862  &      0.030		&    \\
10:24:02.604 &  -57.45:32.26     & O6--7 V			& 16.118	& -0.276  &		    1.579 & 1.661	& 1.543	 &    0.016 &    1.862  &      0.039		&    \\
10:24:02.664 &  -57.45:34.38     & O3--4 V			& 14.554	& -0.282  &		    1.528 & 1.620	& 1.543	 &    0.005 &    1.862  &      0.016		&    \\
10:24:02.789 &  -57.45:30.05     & O8 V				& 15.963	& -0.267  &		    1.593 & 1.667	& 1.543	 &    0.017 &    1.862  &      0.035		&    \\
10:24:03.787 &  -57.44:39.87     & O9.5 V				& 15.349	& -0.259  &		    1.380 & 1.470	& 1.543	 &    0.012 &    1.862  &      0.022		&   b \\
10:24:04.901 &  -57.45:28.43     & O4--5 V			& 14.528	& -0.281  &		    1.606 & 1.688	& 1.511	 &    0.079 &    1.787  &      0.012		&  	b \\

\enddata
\tablecomments{All stars with a determined spectroscopic type from \citet{Alvarez_13} and \citet{Rauw_07,Rauw_11}. All WFPC2 photometry was adopted from \citet{Alvarez_13}. Columns 1--2 list the J2000 coordinates followed by the spectral type. Column 4 gives the WFPC2 F555W magnitude. Columns 5 gives the WFPC2 $F439W-F555W$ theoretical intrinsic colors, derived from \citep[p.23, Tab.1.4,][]{Sparke_07} in combination with the PARSEC ZAMS. Column 6 lists the stellar WFPC2 $F439W-F555W$ colors. Column 7 gives the stellar excess derived from the $UBV$ TCD. Columns 8--11 list the stellar and gas excess $E(B-V)$ and their respective uncertainties derived from our color excess map . Column 12 gives additional information: (a) stars have a proper $F160W$ and $F814W$ photometry in our data and are shown in our CMD (Fig. \ref{fig:F814W-F160W_F814W_4kpc}), (b) individual reddening from the gas extinction map is available.}
\end{deluxetable*}

\subsection{The stellar color excess derived with $UBV$ photometry}
\label{subsec:UBV_TCD}
An independent way of deriving the stellar reddening towards Wd2 is to use $UBV$ two-color diagrams (TCD). Since our own photometric catalog lacks the $U$-band photometry, we used the photometric catalog of \citet{Alvarez_13}, including the spectroscopy derived by \citet{Rauw_07,Rauw_11}, and \citet{Alvarez_13}. We used all stars with a spectral classification of luminosity class V, meaning that they are located on the main-sequence, and have no abnormal line features (see Tab. \ref{tab:comp_spec_data}).

Using the PARSEC ZAMS with the same procedure as described in Subsect. \ref{subsec:excess_transformation}, we were able to obtain the locus of the spectroscopically determined stars on the main-sequence in the $UBV$ TCD. By shifting the individual stars in the TCD to their theoretical loci on the main sequence we could derive the stellar color excess $E(F439W-F555W)_\star$ and $E(F336W-F439W)_\star$. Column 7 of Tab. \ref{tab:comp_spec_data} lists the derived values transformed to the Johnson-Cousins photometric system with the extinction law of \citet{Cardelli_89} (see Eq. \ref{eq:color_excess_transformation}).

The slope of the two-color excess in the TCD gives a ratio of $E(U-B) / E(B-V) = 0.85 \pm 0.033$. Therefore, it is larger than the mean throughout the MW, which amounts to $E(U-B) / E(B-V) = 0.72$ \citep{Fitzgerald_70}. This is consistent with the higher total-to-selective extinction derived in Subsect. \ref{subsec:R_V}. The mean of the color excess of the stars derived via the TCD is $E(B-V)_\star=1.62 \pm 0.113$. The median of the color excess of the stars derived with the color excess map was $E(B-V)_\star=1.56$. Comparing the individual values of $E(B-V)_\star$ derived with the TCD with those of the color excess map reveals a mean difference of $\Delta E(B-V)_\star = 0.087 \pm 0.129$. We can say, within the errors, that our method of deriving the pixel-to-pixel gas excess map and transferring it to an excess map of the stars was successful and we can confidently use this map to deredden our photometry of the cluster members.

\section{The Color-Magnitude Diagram of Wd2}
\label{sec:CMD}

\subsection{The morphology of the CMD}
\label{subsec:CMD_morphology}
\begin{figure}[htb]
\resizebox{\hsize}{!}{\includegraphics{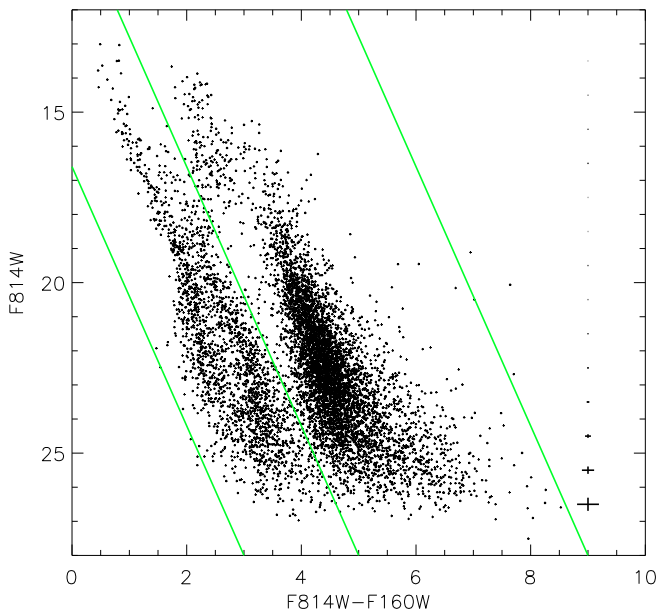}}
\caption{The $F814W$ vs. $(F814W-F160W)$ CMD of the Wd2 region. The plot contains 7\,697 objects. On the right representative photometric error bars are shown in bins of 1~mag. The green lines represent the separation between the two parallel sequences of likely foreground stars (blue sequence) and point sources likely associated with Wd2.}
\label{fig:F814W-F160W_F814W}
\end{figure}

In order to understand the stellar populations in our dataset we plotted the $F814W-F160W$ vs. $F814W$ color-magnitude diagram (CMD) of our catalog (see Fig. \ref{fig:F814W-F160W_F814W}). This CMD contains 7\,697 objects in total. Looking at the CMD it immediately becomes clear that we observe a composite population. There are two parallel sequences: a blue one at $F814W-F160W \sim 2$--4~mag, and a redder one, at $F814W-F160W \sim 4$--8~mag (with approximate boundaries represented by the green lines in Fig. \ref{fig:F814W-F160W_F814W}). The bluer population seems to consist of two branches, probably due to stars at different distances along the line of sight through the Galactic disk. The brightest objects can be seen at $F814W \approxeq 14$~mag.

The bluer sequence of the two clearly separated populations can be explained with the foreground population of the Galactic disk towards Wd2 (see Sect. \ref{sec:physical_parameters}). This population is located closer to us and hence less reddened. This makes those stars appear bluer in the CMD (see the comparison to the Besan\c{c}on model \citep{Robin_03} in Fig. \ref{fig:F814W-F160W_F814W_model}). The red sequence of the CMD is most likely the cluster population of Wd2 (see Sect. \ref{subsec:cluster_members}).

Our data suffer from saturation of the very bright objects (see Sect. \ref{sec:catalog_creation}), and for this reason we do "lose" stars in the CMD at the bright end. Due to the detection method in the WFC3-IR filters (see Sect. \ref{sec:observations}) saturation occurs only in the optical filters. Comparing our objects with detections in at least two filters with more shallow photometric studies from the literature \citep{Alvarez_13} we conclude that we miss 16 luminous stars, at the bright end of our photometry.

\subsection{Selection of likely RCW~49 member stars}
\label{subsec:cluster_members}
In order to study the properties of Wd2 and the spatial morphology of the stellar content we need to extract a sample of probable RCW~49 member stars. Therefore, we need to clean the catalog from foreground objects (see Sect. \ref{sec:CMD}).

\begin{figure}[htb]
\resizebox{\hsize}{!}{\includegraphics{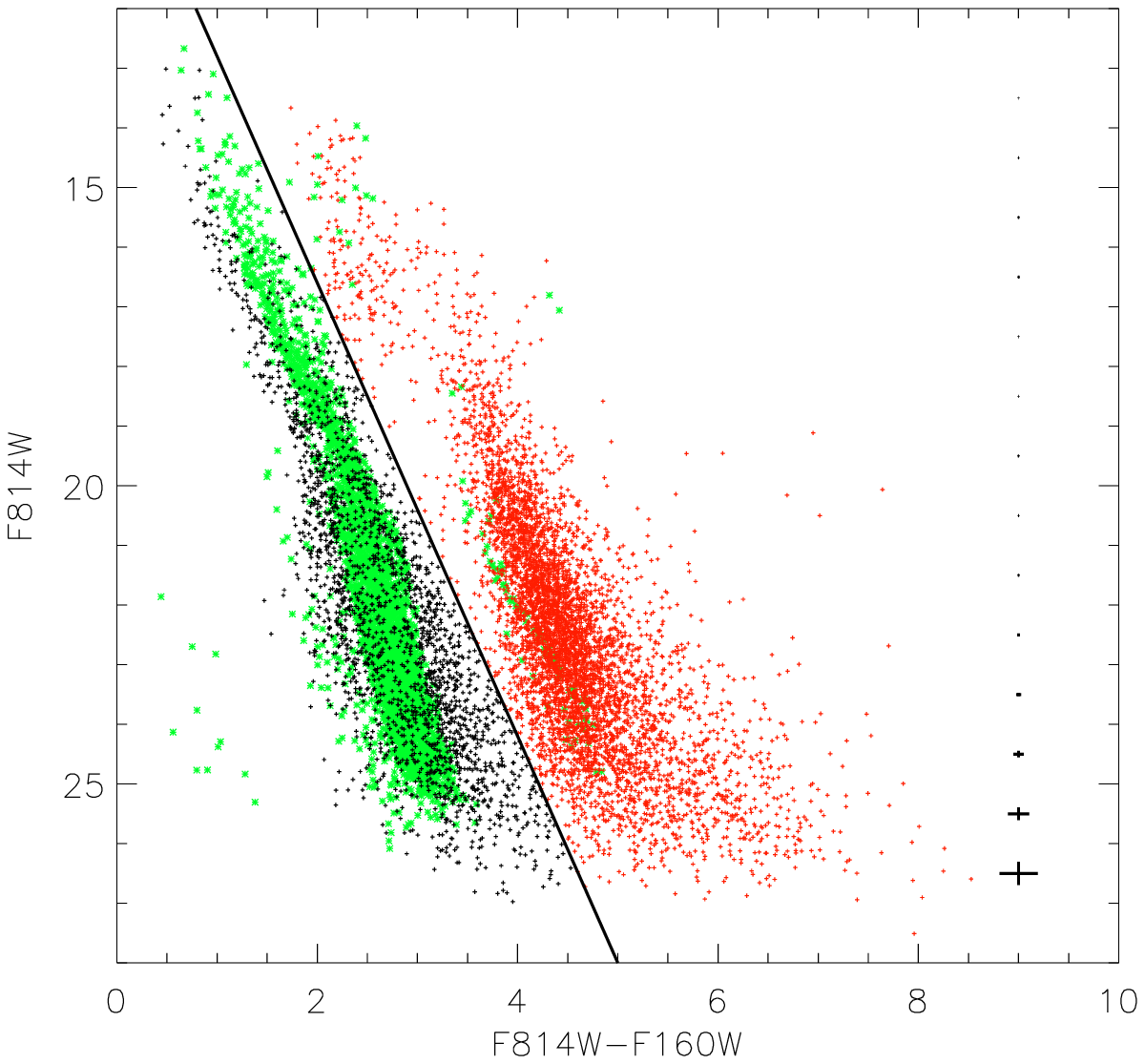}}
\caption{The $F814W$ vs. $(F814W-F160W)$ CMD of the 7\,697 point sources from the Wd2 photometric catalog. The red points represent likely members of the RCW~49 region (including cluster members). The black sources are probable foreground stars. The green points indicate the location of the synthetic objects from the Besan\c{c}on model in the direction of Wd2 up to the cluster distance of 4.16~kpc falling within a FOV corresponding in size to our survey area ($\sim 21$~arcmin$^2$). Their superposition with the black points suggests that the latter are less reddened foreground stars. The black solid line marks our arbitrary separation between cluster and field stars.}
\label{fig:F814W-F160W_F814W_model}
\end{figure}

The RCW~49 member selection was performed using the $F814W-F160W$ vs. $F814W$ CMD of our catalog (see Fig. \ref{fig:F814W-F160W_F814W_model}) which contains a total number of 7\,697 objects (see Sect. \ref{sec:CMD}). As a first step, we divided the data in likely foreground population stars (5\,604) (marked by black points) and the likely RCW~49 member stars (2\,093), indicated by red points (see Fig. \ref{fig:F814W-F160W_F814W_model}). Distinguishing these two populations by defining a diagonal line was possible due to the clear separation of the two sequences and even at the faint end of the CMD the photometric uncertainties are still small enough to clearly separate the sequence formed by unrelated Galactic disk objects and RCW~49 members.

To test whether the black objects are indeed contaminant stars of the Galactic plane, we simulated the MW disk in the direction of Wd2 with the \texttt{Model of stellar population synthesis of the Galaxy}, the Besan\c{c}on model \citep{Robin_03}. This model adopts the extinction law of \citet{Mathis_90}. It lacks the spiral arm structure \citep{Robin_03}, but since we are just sampling one line of sight, this should not be a problem. We simulated the color-magnitude locus of the expected stars in the Galactic plane up to the cluster distance of 4.16~kpc across a FOV corresponding in size to our survey area ($\sim 21$~arcmin$^2$). The simulated stellar distribution is visualized by the green points in Fig. \ref{fig:F814W-F160W_F814W_model}. We transformed the model colors and bands, which are by default in the Johnson-Cousins photometric system \citep{Johnson_53}, into the HST photometric system with the \textsc{synphot/calcphot} routine\footnote{Synphot is a product of the Space Telescope Science Institute, which is operated by AURA for NASA.}\citep{Synphot} for consistency with the data. It looks like the green points are centered between the two slightly separated populations of the foreground, most likely originating from a non-uniformly distributed ISM in the line of sight towards Wd2. The green points do indeed nicely overlap with the blue diagonal sequence of suspected foreground stars, supporting our assumption. Since we do not know the distribution of the foreground reddening we cannot reproduce the exactly observed distribution of the foreground populations.

\section{The physical parameters of Wd2}
\label{sec:physical_parameters}

Using our high resolution dataset in combination with the 2D color excess map we are able to determine the physical properties of Wd2.

\subsection{The total-to-selective extinction towards the direction of Wd2}
\label{subsec:R_V}
In order to correct for the high differential reddening, visible in the map of the color excess $E(B-V)_g$ (see Fig. \ref{fig:red_map}.), we need to get a better insight in the total-to-selective extinction ratio $R_V$:

\begin{equation}
\label{eq:R_V}
R_V=\frac{A_V}{E(B-V)},
\end{equation}

where $A_V$ represents the total extinction in the visible.

\begin{figure}[htb]
\resizebox{\hsize}{!}{\includegraphics{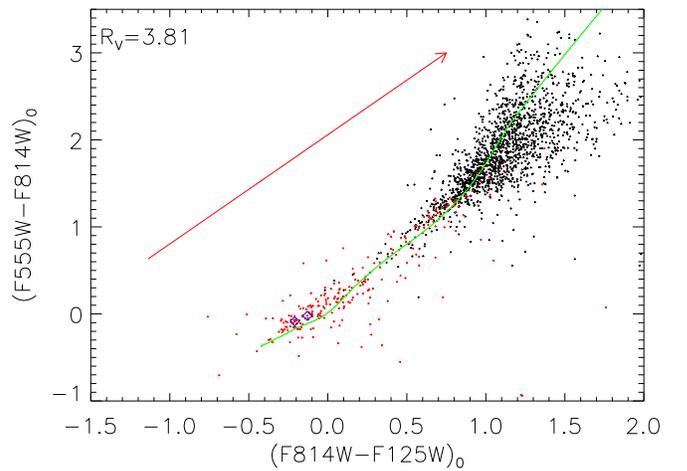}}
\caption{The $F555W-F814W$ vs. $F814W-F125W$ TCD of the Wd2 photometric catalog. The green line represents the PARSEC ZAMS. We show in red the reddening vector for $E(B-V)_\star=1.55$~mag and $R_V=3.81$. The red points represent the likely main-sequence stars (all RCW~49 members brighter than $F814W>18$~mag). The discrepancies at the red end originate from the rich PMS population appearing redder due to the persisting dusty envelopes around the stars. The blue diamonds mark the stars with spectral types from \citep{Alvarez_13}.}
\label{fig:F555W-F814W_F814W-F125W}
\end{figure}

\begin{figure}[htb]
\resizebox{\hsize}{!}{\includegraphics{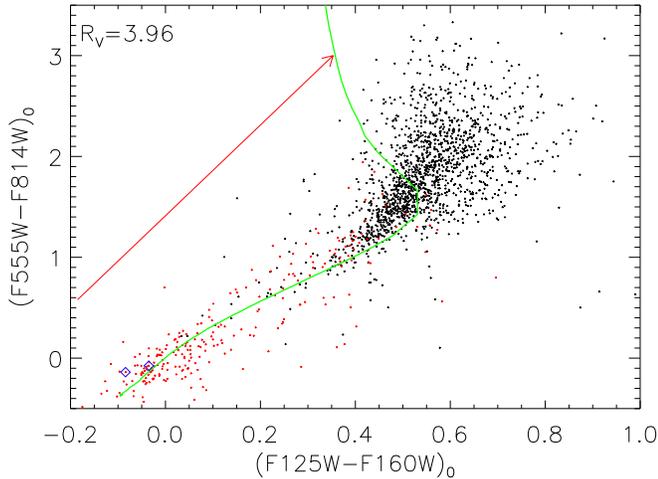}}
\caption{The $F555W-F814W$ vs. $F125W-F160W$ TCD of the Wd2 photometric catalog. The green line represents the PARSEC ZAMS. In red we show the reddening vector for $E(B-V)_\star=1.55$~mag and $R_V=3.96$, shortened by a factor of 4 for plotting reasons. The red points represent the likely main-sequence stars (all RCW~49 members brighter than $F814W>18$~mag). The discrepancy at the red end originate from the rich PMS population appearing redder due to the persisting dusty envelopes around the stars. The blue diamonds mark the stars with spectral types from \citep{Alvarez_13}.}
\label{fig:F555W-F814W_F125W-F160W}
\end{figure}

Since the distance modulus for Wd2 is not yet fixed, we used the TCD to derive a value for $R_V$, since TCDs have the advantage of being distance independent. To derive $R_V$, we had to find the best fit of the ZAMS to the likely main-sequence stars of our photometric catalog. Therefore, we used all stars brighter than $F814W>18$~mag of the RCW~49 members (red points in Fig. \ref{fig:F555W-F814W_F814W-F125W} and Fig.  \ref{fig:F555W-F814W_F125W-F160W} selected through the CMD of Fig.\ref{fig:F814W-F160W_F814W_model}). We used three different TCDs: $F555W-F814W$ vs. $F814W-F125W$, $F555W-F814W$ vs. $F125W-F160W$, and $(F439W-F555W)_{WFPC2}$ vs. $(F555W-F814W)_{WFPC2}$. Here again, we also used the WFPC2 data from the photometric catalog of \citet{Alvarez_13}. The PARSEC ZAMS were used for Solar metallicity.

In order to deredden the stars onto the locus of the ZAMS in the TCDs we used the extinction law of \citet{Cardelli_89}:

\begin{equation}
\label{eq:reddening_transformation}
A_\lambda=\left( \frac{A_\lambda}{A_V}\right) \cdot E(B-V)_\star \cdot R_V.
\end{equation}

$A_\lambda / A_V$ is also $R_V$-dependent:

\begin{equation}
\label{eq:Alambda_AV}
\left( \frac{A_\lambda}{A_V}\right)=a_\lambda +\frac{b_\lambda}{R_V}.
\end{equation}

\begin{figure}[htb]
\resizebox{\hsize}{!}{\includegraphics{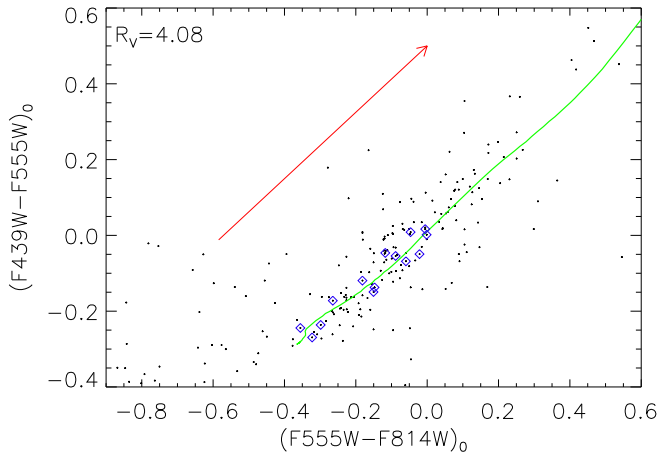}}
\caption{The $(F439W-F555W)_{WFPC2}$ vs. $(F555W-F814W)_{WFPC2}$ TCD of the WFPC2 photometric catalog of \citet{Alvarez_13}. The green line represents the PARSEC ZAMS. The reddening vector for $E(B-V)_\star=1.55$~mag and $R_V=4.08$ is shown in red. The blue diamonds mark the stars with spectral types from \citep{Alvarez_13}.}
\label{fig:F439W-F555W_F555W-F814W}
\end{figure}

The parameters $a$ and $b$ are wavelength dependent and given in Tab. \ref{tab:color_excess_transformation}. Combining Eq. \ref{eq:reddening_transformation} and Eq. \ref{eq:Alambda_AV} we can calculate the absolute extinction $A_\lambda$ in dependence of $R_V$. In order to get the best fit for the three TCDs we needed to apply values of $R_V=3.81$ for $F555W-F814W$ vs. $F814W-F125W$ (see Fig. \ref{fig:F555W-F814W_F814W-F125W}), $R_V=3.96$ for $F555W-F814W$ vs. $F125W-F160W$ (see Fig. \ref{fig:F555W-F814W_F125W-F160W}), and $R_V=4.08$ for $(F439W-F555W)_{WFPC2}$ vs. $(F555W-F814W)_{WFPC2}$ (see Fig. \ref{fig:F439W-F555W_F555W-F814W}). In the corresponding Figures \ref{fig:F555W-F814W_F814W-F125W}, \ref{fig:F555W-F814W_F125W-F160W}, and \ref{fig:F439W-F555W_F555W-F814W} we show the stellar population including the ZAMS and the reddening vector. The red points represent the likely main-sequence stars, the blue diamonds represent the objects for which spectroscopy is available. The discrepancies at the red end of the IR filters in TCD to the remaining objects (black points) originate from the very numerous PMS population appearing redder due to remaining circumstellar dusty envelopes. Due to the high number of spectroscopically observed stars in Fig. \ref{fig:F439W-F555W_F555W-F814W}, we only used those objects for fitting. The ZAMS fitting gives us a mean value of $R_V=3.95 \pm 0.135$. Spectroscopic observations of the most massive stars \citep{Alvarez_13,Rauw_07,Rauw_11} and a photometric study of \citet{Hur_15} show that Wd2 has a total to selective extinction in the visual ($R_V=A_V/E(B-V)$) of about 3.64--3.85, what agrees with our value within the uncertainties.

\subsection{The extinction correction of our cluster CMD}
\label{subsec:ext_cor}

The high resolution color excess map of the gas $E(B-V)_g$, was transformed to the stellar color excess $E(B-V)_\star$ (see Subsect. \ref{subsec:excess_transformation}) and used to correct the CMD for differential reddening (see Sec. \ref{subsec:properties_red_map}). In other words, we used the stellar color excess at the position of the stars to subtract the foreground reddening. For the color excess $E(B-V)_\star$ at each stellar position, the median $E(B-V)_\star$ within three sigma of the mean PSF was calculated and then translated into the total extinction $A_\lambda$ at the pivot wavelength $\lambda$ of the HST filters, using again the extinction law of \citet{Cardelli_89} (see Eq. \ref{eq:reddening_transformation}). $A_\lambda / A_V$ is given as $0.62303$ for $F814W$ and $0.21756$ for $F160W$.

\begin{figure*}[htb]
\plottwo{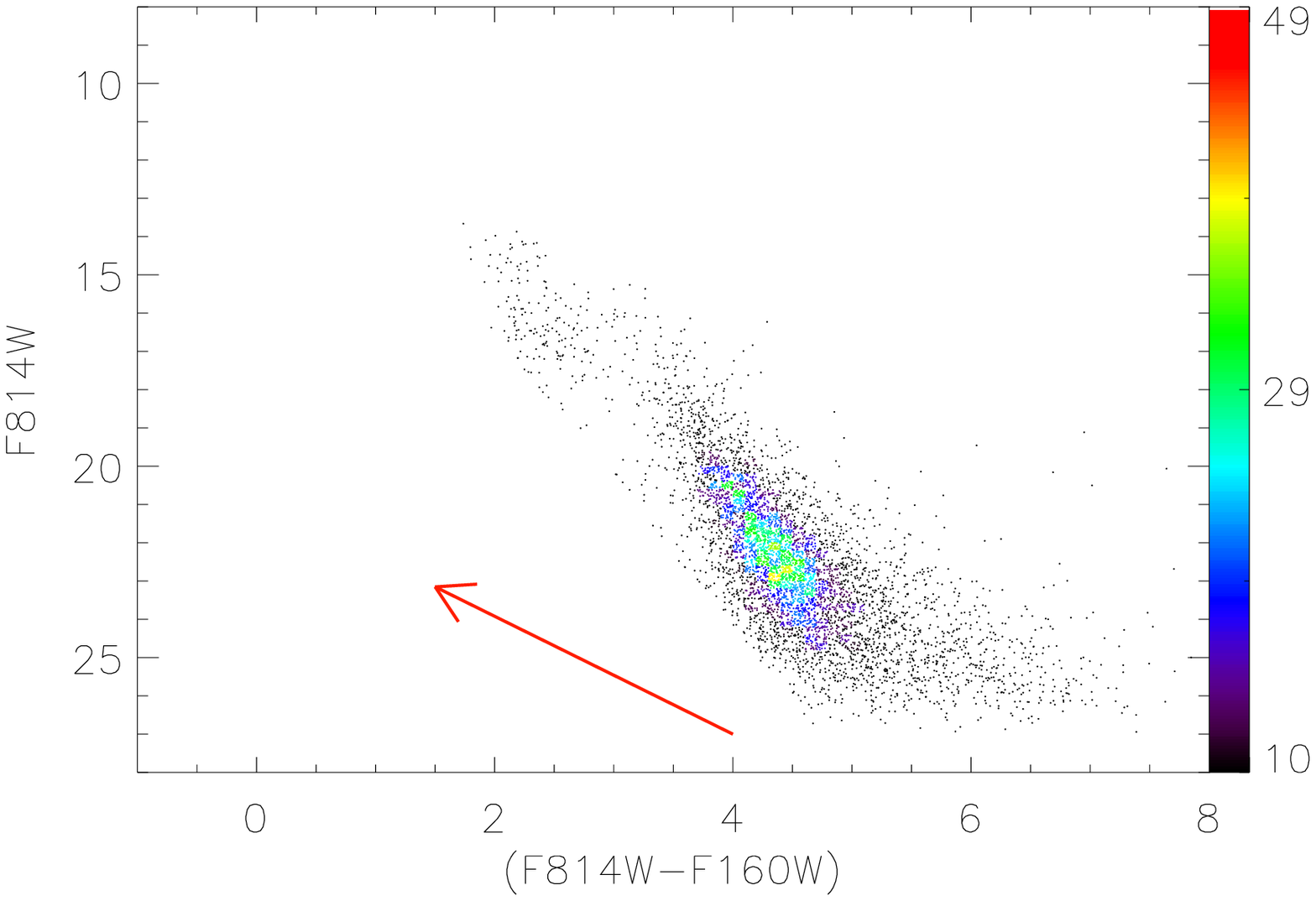}{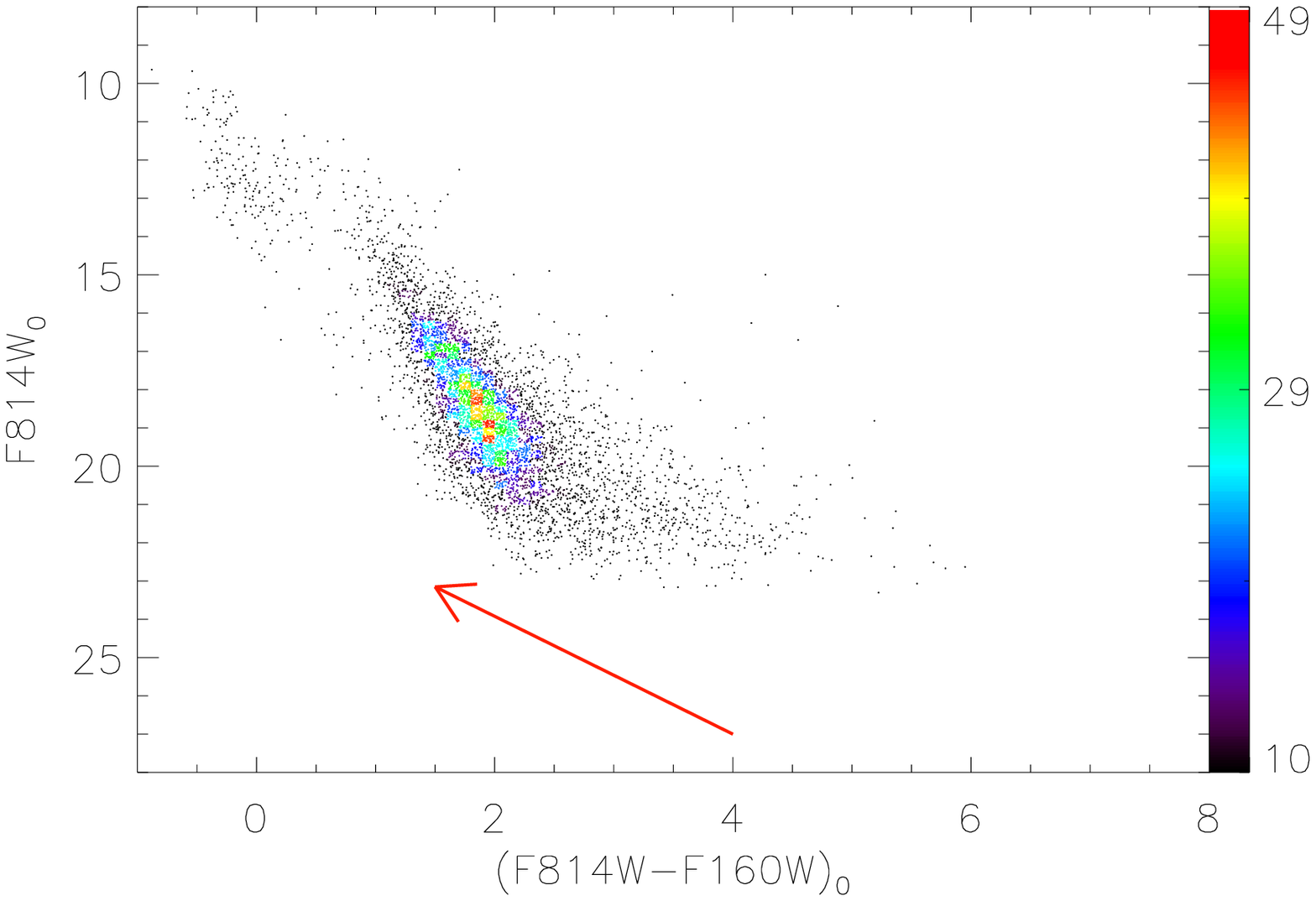}
\caption{\textbf{Left: }The observed $F814W$ vs. $(F814W-F160W)$ CMD of the likely cluster members. \textbf{Right: }The $F814W$ vs. $(F814W-F160W)$ CMD of the of the individually dereddened cluster stars. The red arrow marks the reddening vector corresponding to the median $E(B-V)_\star=1.55$~mag with $R_V=3.95$. The colorbar represents the number of objects per bin (binsize: $0.1 \time 0.2$~mag). The range in color is the same in both plots.}
\label{fig:F814W-F160W_F814W_red_dered}
\end{figure*}

In Fig. \ref{fig:F814W-F160W_F814W_red_dered} we show the CMD of the Wd2 members as observed (left) as well as the individually dereddened CMD (right). The red arrow shows the overall median reddening of $E(B-V)_\star=1.55$ with $R_V=3.95$ (see Subsect. \ref{subsec:R_V} for determining $R_V$). We should note here that objects lying in regions where the color excess map needed to be interpolated or substituted by the median (see Sect. \ref{subsec:properties_red_map}), the uncertainty of the dereddening is bound to the uncertainty of the estimation of the values of the saturated pixels. Consequently, individual dereddening means using the individual stellar excess for each star and the extrapolated/substituted value for the stars where the color-excess map has corrupt pixels. As one can see from the color representation of the density in the plot, a tightening of the CMD was achieved. Nevertheless, the spread cannot be reduced to less than the intrinsic age spread of $\sim 2$~Myr. Also effects like rotation and and binarity play a role. PMS stars have their own circumstellar effects, in addition to the general differential reddening, causing broadening of the CMD.

\subsection{The distance and age}
\label{subsec:dist_age}
The distance to Wd2 is still subject of debate in the literature. Published values range from 2.8~kpc \citep{Rauw_07,Ascenso_07,Carraro_13}, to 4.16~kpc \citep{Alvarez_13}, 5.7~kpc \citep{Piatti_98}, 6.4~kpc \citep{Carraro_04} to 8~kpc \citep{Rauw_07,Rauw_11}. Discrepancies also exist in the published ages: an upper age limit of 3~Myr was inferred for the whole cluster and 2~Myr for the core by \citet{Ascenso_07} and \citet{Carraro_13}.

In this paper, we will attempt to derive tighter constraints on distance and age from our high resolution, high sensitivity multi-wavelength observation.

In order to determine the distance and age of Wd2 we used the method of over-plotting isochrones to our $F814W$ vs. $(F814W-F160W)$ CMD. We especially focused on the turn-on, since here the isochrones are not degenerate in age and most distinguishable between each other, compared to the main sequence. We used PARSEC isochrones for 0.5--2.0~Myr and Solar metallicity.

\begin{figure}[htb]
\resizebox{\hsize}{!}{\includegraphics{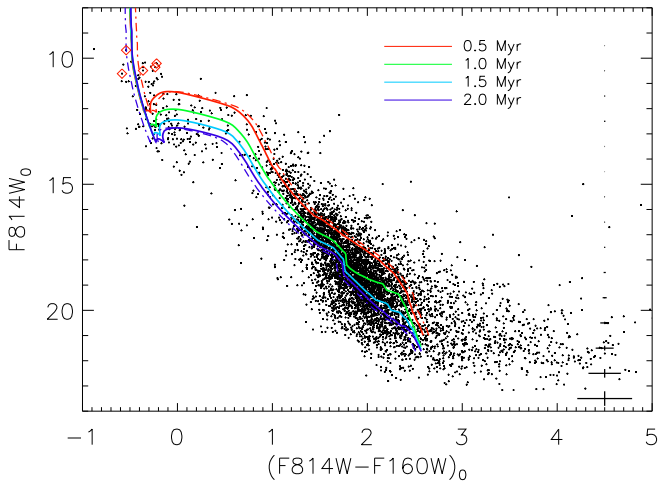}}
\caption{The extinction corrected $F814W$ vs. $(F814W-F160W)$ CMD of the 5\,418 point sources left the Wd2 member selection. The colored lines represent the PARSEC PMS isochrones for 0.5--2.0~Myrs best fitting the turn-on of the main sequence. The error bars represent the typical photometric uncertainties in one magnitude bins. We applied an $R_V=3.95$ with a distance of 4.16~kpc. The red diamonds mark the stars for which spectroscopic data are available (see Tab. \ref{tab:comp_spec_data}). The dashed-dotted lines represent the maximum possible shifts of the isochrones for $\Delta R_V \pm 0.135$ assuming a constant luminosity of the stars.}
\label{fig:F814W-F160W_F814W_4kpc}
\end{figure}

In order to determine the best distance, we used the already dereddened CMD (see Fig. \ref{fig:F814W-F160W_F814W_4kpc}, right panel). We corrected the chosen PARSEC isochrones \citep{Bressan_12} for the distance modulus. We applied different values for the distance and assessed the best fit to both the locus of the turn-on region and the PMS population. This corresponded to a distance of $d=4$~kpc and is in good agreement with the spectroscopic observations of \citet{Alvarez_13}. Their distance is based on the physical constraints derived from the spectral types of the most massive stars in Wd2. The method they adopted is more accurate than our empirical fit, and, therefore, we will assume for all the subsequent calculations their value of $d=4.16$~kpc.

Taking into account the magnitude range of the turn-on we can conclude that the age for the whole cluster is in the range of 0.5--2.0~Myr. We provide a more detailed and spatially differentiated analysis of the age in Sect. \ref{sec:two_clumps}. The uncertainty in the total-to-selective extinction $\Delta R_V \pm 0.135$ results in an age uncertainty of $\sim 0.5$~Myr (see dashed-dotted lines in Fig. \ref{fig:F814W-F160W_F814W_4kpc}) assuming that the luminosity of the stars is constant. The impact on the distance is $\pm 0.01$~kpc and, therefore, negligible.

\section{The two clumps of Westerlund 2}
\label{sec:two_clumps}

\subsection{The spatial distribution of the stellar content}
\label{subsec:spatial_dist}

As can be seen from Fig. \ref{fig:RGB_JIV}, our observations cover an area $\sim4$~arcmin wide, which, assuming a distance of 4.16~kpc, corresponds to a linear scale of 4.8~pc. This region covers the Wd2 cluster and the surrounding clouds, and enables us to analyze the spatial distribution of the stars. Due to the very complex cloud structure and the resulting variable reddening (see Sect. \ref{sec:reddening_map}) as well as the different degrees of crowding, we expect that the completeness of our dataset differs as a function of position. In order to avoid the bias due to incompleteness effects we applied luminosity cuts based on the histogram of the magnitude distribution (see Fig. \ref{fig:completeness}). In other words, we limited our analysis to the range where the catalog is mostly complete. We used all objects with $F814W<22.75$~mag and $F160W<18.00$~mag. This leaves us with 2\,610 point sources. A detailed assessment of the incompleteness through artificial star experiments will be performed and published in a subsequent paper.

\begin{figure*}[htb]
\resizebox{\hsize}{!}{\includegraphics{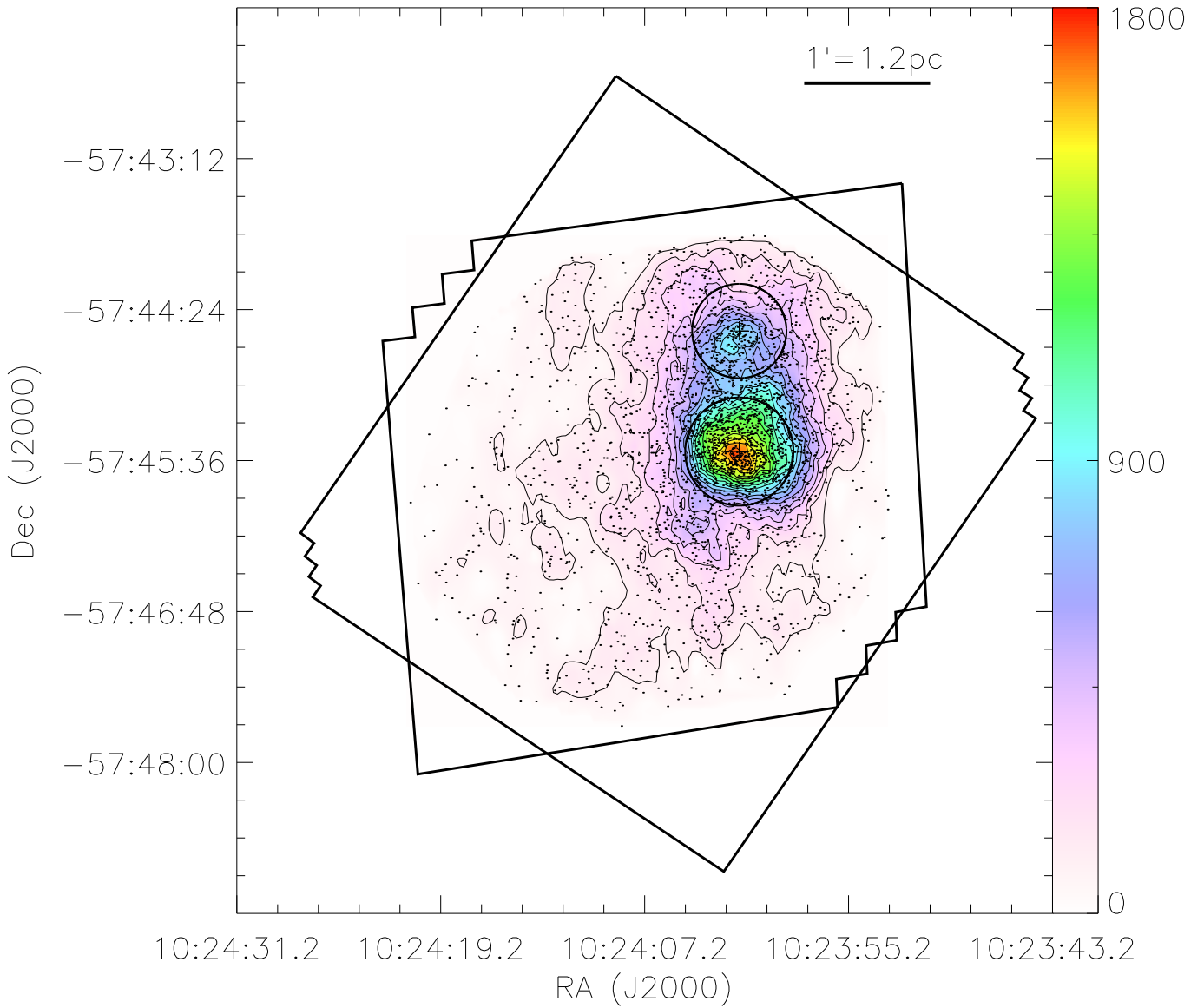}}
\caption{The surface number density of the Wd2 stellar content. Each dot represents one single star. The color bar indicates the density in number arcmin$^{-2}$. The contours show the density in 15 linear steps. The outer boundaries of our survey are indicated by thick black lines. The approximate radii of Wd2 ($r_{Wd2}=24.10''$) and the northern clump ($r_{nc}=18.96''$) determined by the FWHM of a Gaussian distribution are shown by two large circles.}
\label{fig:density_map}
\end{figure*}

Using this reduced sample we created a 2D plot of the stellar density, taking into account all objects within a radius of $r<10.8''$ around each star. This area was chosen such that there were always enough objects within the area and the resolution was high enough to resolve the spatial structure of the stellar content. The result of this approach can be seen in Fig. \ref{fig:density_map}. Each dot represents one star. Fifteen linear spaced lines representing isodensity contours are over-plotted. The color indicates the number density specified by the color bar [number/arcmin$^2$].

In the resulting surface density plot the cluster appears divided into two separate clumps; the main cluster of Wd2 and a smaller clump to the north (northern clump). This structure was also pointed out and named northern clump by \citet{Hur_15}. \citet{Alvarez_13} also noticed a "a secondary concentration of stars $45''$ to the north".

In order to investigate the properties of and possible differences between the two clumps we defined a circular area around each clump containing stars that we considered to belong to each clump. We removed the uniform distribution of stars, visible outside the cluster, by determining a threshold density. This density was determined in a control field well outside the central cluster. All objects $3\sigma$ above this stellar density were included.

\begin{figure*}[htb]
\plottwo{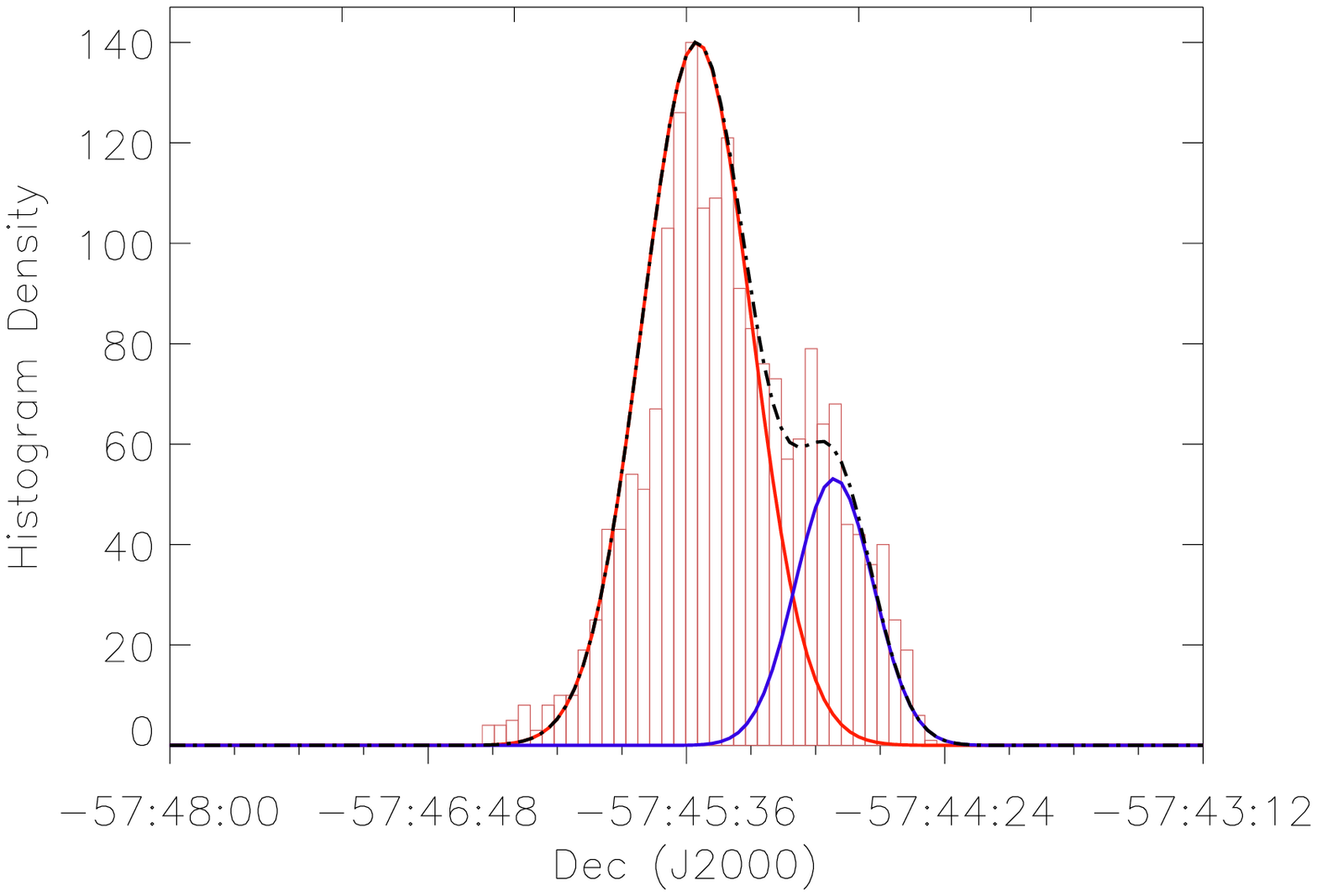}{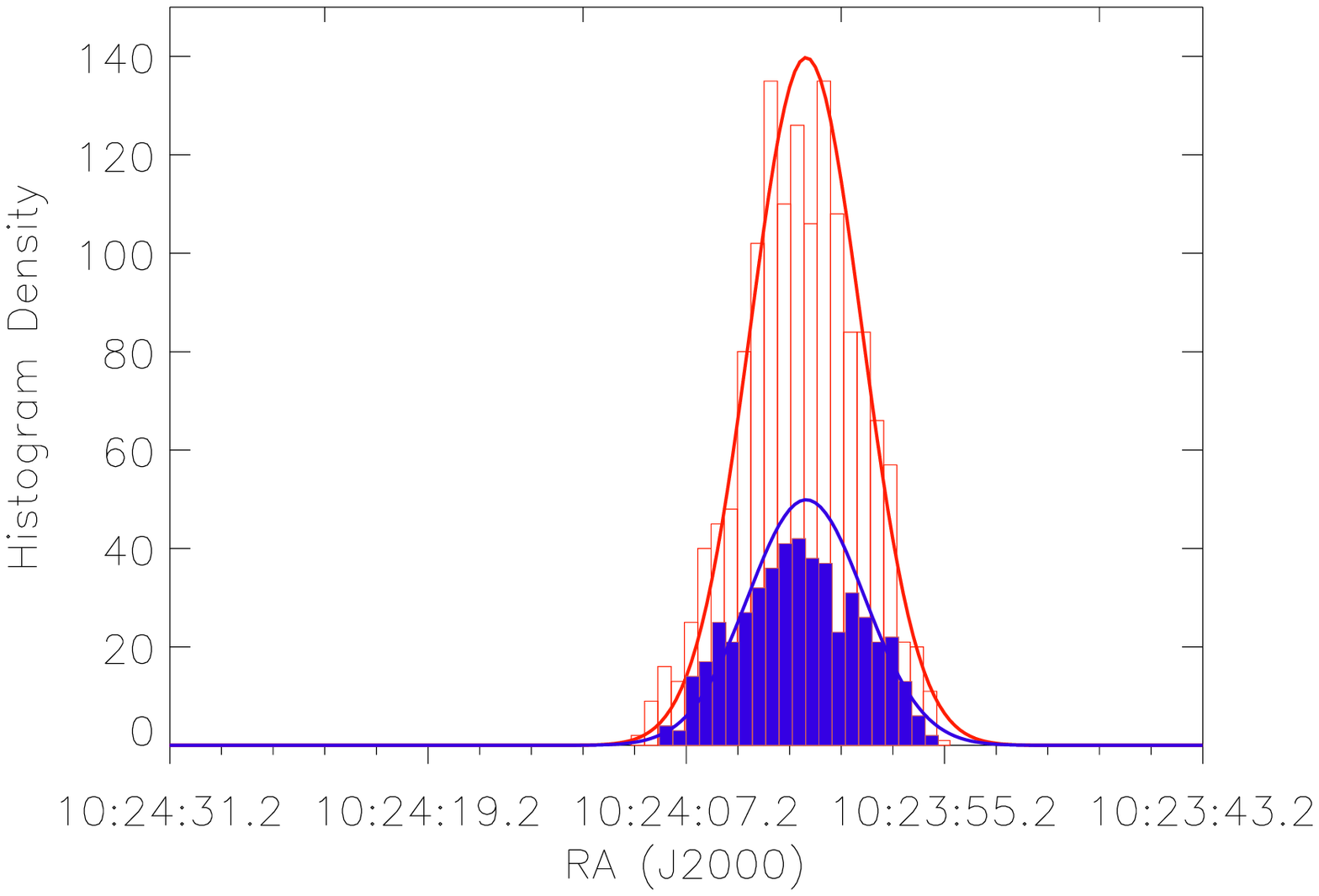}
\caption{Histograms of the number of stars as a function of spatial position (expressed in right ascension and declination) with a binsize of 5~arcsec. \textbf{Left:} Histogram in the declination direction with a superimposed curve fitting two Gaussians (dashed-dotted line). The red curve is a fit of the maximum of the histogram distribution and marks the center of the main concentration of Wd2 which we find to be at a declination of $-57^\circ45'31''7$ and to have a FWHM of $26.83''$. The blue curve fits the northern clump with a center coordinate of $-57^h44^m34^s2$ and a FWHM of $19.17''$. \textbf{Right:} Histograms along the right ascension direction for the main body of Wd2 (red) and the northern clump (blue). The red curve fits the main body of Wd2 with a center coordinate of 10:24:01.63 and a FWHM of $24.76''$. The blue curve fits the northern clump with a center coordinate of 10:24:01.63 and a FWHM of $24.76''$.}
\label{fig:histograms}
\end{figure*}

We then plotted histograms in right ascension and declination of all remaining objects (Fig. \ref{fig:histograms}). Fitting Gaussians to the distribution and determining the FWHM allowed us to estimate the extent of the area where the two clumps dominate. The fits were always made under the condition that the peaks in right ascension and declination for the same clump needed to be similarly high in order to be consistent.

For the declination direction we used a combination of two Gaussians (dashed-dotted-line) (see left of Fig.\ref{fig:histograms}). The red curve fits the maximum of the histogram distribution with a center coordinate of $-57^\circ45'31''7$ and a FWHM of $26.83''$. This coordinate marks the locus of the main body of Wd2. The blue curve fits the northern clump with a center coordinate of $-57^\circ44'34''2$ and a FWHM of $19.17''$.

For the right ascension direction, we needed to subdivide the data since the two clumps are located at very similar right ascensions. Therefore, we divided the catalog at Dec=$-57^\circ44'51''0$. We plotted the histograms for each part (see right side of Fig. \ref{fig:histograms}). Stars associated with the main body of Wd2 are plotted in red and those with the northern clump in blue. The red curve fits the main body of Wd2 with a center coordinate of RA$=10^h24^m01^s.63$ and a FWHM of $24.76''$. The blue curve fits the northern clump with a center coordinate of RA$=10^h24^m01^s.63$ and a FWHM of $24.76''$. The Gaussian fit of the northern clump is higher than the actual distribution due to the fact that for both declination and right ascension the peaks needed to be of similar height. Since we do not have any completeness information, yet, we assumed a circular shape for the clumps. We will be able to determine more accurately the spatial structure of those clumps after having assessed the completeness of our photometry. This will be included in a subsequent paper.

We calculated the radii of the clumps as the mean of the radii along the declination and right ascension directions, and found a radius of $r_{\rm{Wd2}}=24.10''$ ($\sim 0.5$~pc) for the stars associated with the main body of Wd2 and a radius of $r_{\rm{nc}}=18.96''$ ($\sim 0.4$~pc) for the northern clump. These radii are plotted as circles in Fig.\ref{fig:density_map}. These radii merely serve to select stars that mainly belong to one of the two clumps. A thorough analysis of the density profiles (including incompleteness corrections) are deferred to a subsequent paper.

\subsection{The properties of the two clumps}

\subsubsection{The morphology}

\begin{figure*}[htb]
\resizebox{\hsize}{!}{\includegraphics{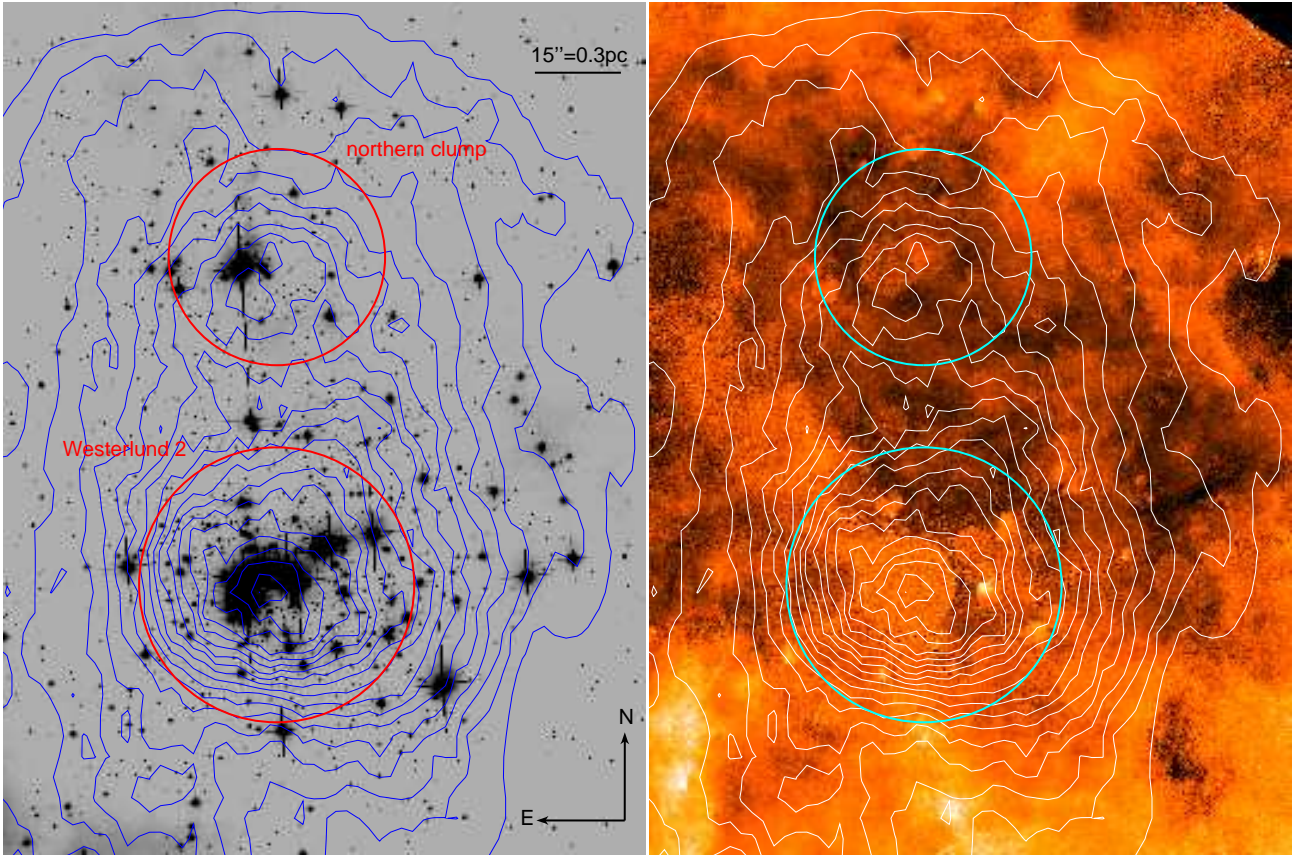}}
\caption{\textbf{Left: } Gray-scale image of the main body of Wd2 and the northern clump in the $F814W$ band. \textbf{Right: }Reddening map shown on the same spatial scale and the same intensity scale as Fig \ref{fig:red_map}: Both panels show the same central areas. The stellar density contours are over-plotted as black (left) and white (right) lines (see Fig. \ref{fig:density_map}). The circles denote the areas dominated by stars belonging to the main body of Wd2 and to the northern clump, respectively.}
\label{fig:reddening+I_band}
\end{figure*}

In Fig. \ref{fig:reddening+I_band} one can see a gray-scale image of the central region containing the two clumps in the $F814W$ band (left panel) and the reddening map of the same area (right panel). The stellar density contour lines from Fig. \ref{fig:density_map} and the circle marking the central regions of the two clumps are over-plotted. Our spatial analysis (see Sect. \ref{subsec:spatial_dist}) shows that the surface density of the stellar content is higher in the clumps than in the surrounding area. The reddening map shows an extended dust filament superimposed on the northern clump. The de-reddened CMDs of both clumps (see Fig. \ref{fig:clump_CMDs}) are not shifted in color which leads to the conclusion that this dust filament is located in front of the cluster.  In the main body of Wd2 the contour lines follow the gas excess towards the North-West, so it looks like the stars there already cleared some of the gas. Towards the South and South-East the gas excess increases and the number density of the stellar content decreases quite rapidly.

\subsubsection{The luminous stellar population}

Both clumps of Wd2 contain multiple OB stars. Most of them are located in the main body of Wd2. All OB stars with available spectroscopic data are saturated in our data.

There are 3 stars with spectroscopic data available from \citet{Rauw_07,Rauw_11}, and \citet{Alvarez_13} within the northern clump and 19 within the main body Wd2. The northern clump contains a B1 V, O5 V-III, and an O9.5 V star. The 19 stars in the main body of Wd2 are all O-type stars with a spectral type ranging from O9.5 V to O3--4 V.

\subsubsection{The color magnitude diagrams}

\begin{figure*}[htb]
\plottwo{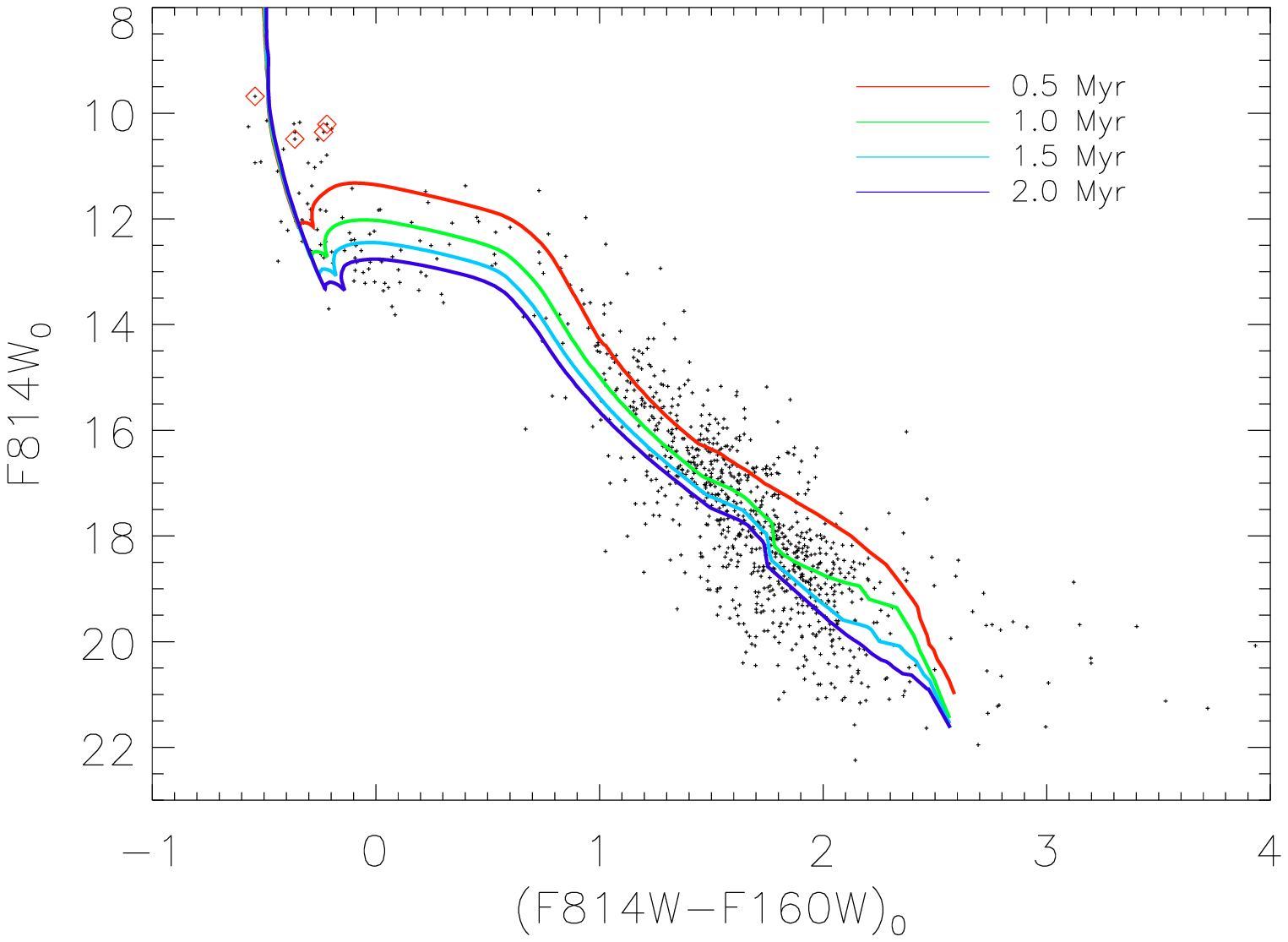}{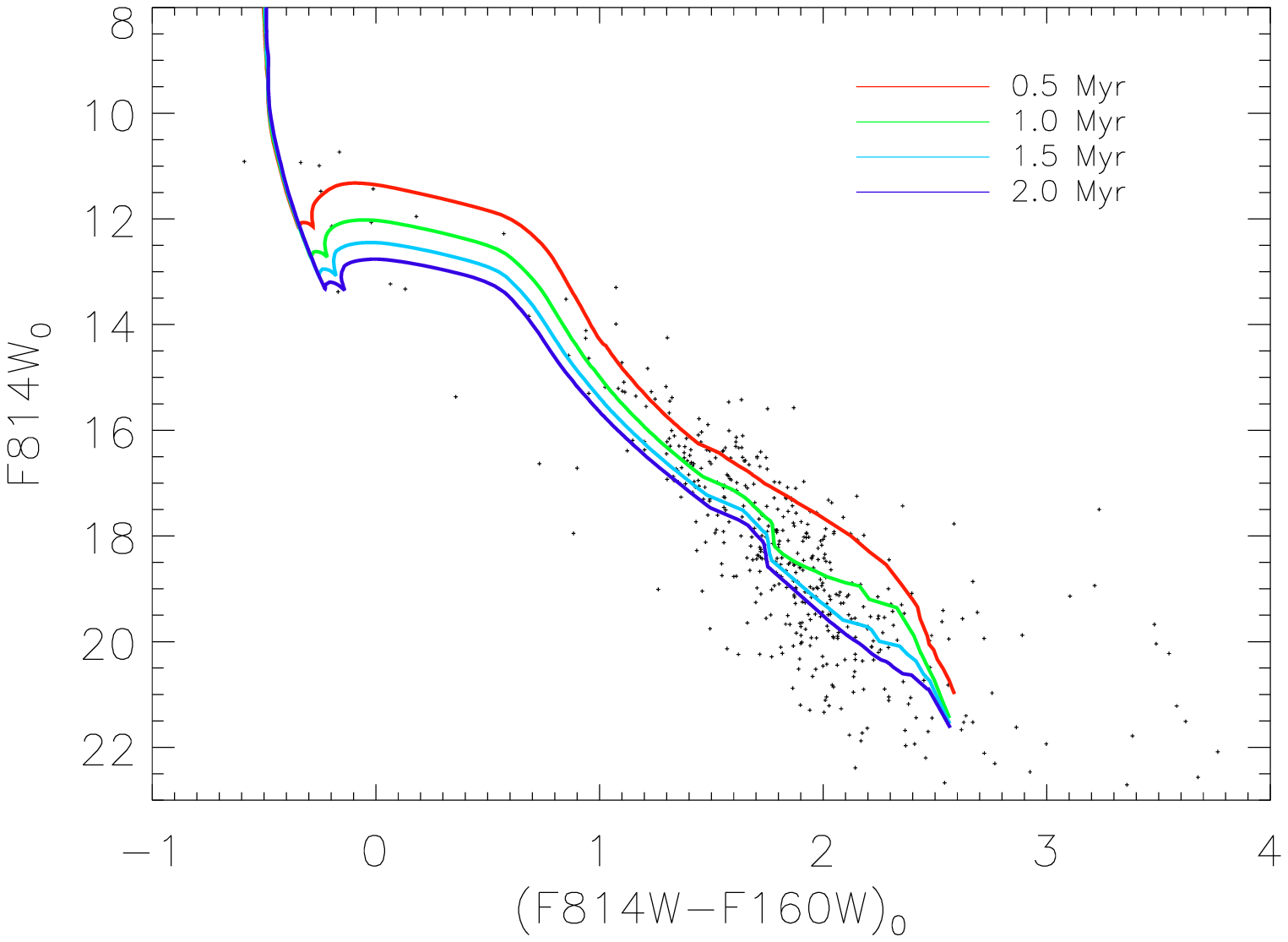}
\caption{The $F814W$ vs. $(F814W-F160W)$ CMDs of Wd2 (left, 969 objects) and of the northern clump (right, 469 objects). The colored lines represent the PARSEC PMS isochrones for 0.5--2.0~Myr fitting the turn-on of the main sequence. The red diamonds in the left panel represent those objects for which spectroscopic data is available (see Tab. \ref{tab:comp_spec_data})}
\label{fig:clump_CMDs}
\end{figure*}

In the next step we take a look at the dereddened CMDs (see Fig.\ref{fig:clump_CMDs}) of the main body of Wd2 and the northern clump. We are assuming the two clumps to be at the same distance. We again use the whole PMS population, taking into account all stars in the respective areas. This yields 969 objects for the main body of Wd2 (left panel of Fig. \ref{fig:clump_CMDs}) and 474 for the northern clump (right panel of Fig. \ref{fig:clump_CMDs}). PARSEC isochrones for ages of 0.5--2.0~Myr are overplotted.

\begin{figure}[htb]
\resizebox{\hsize}{!}{\includegraphics{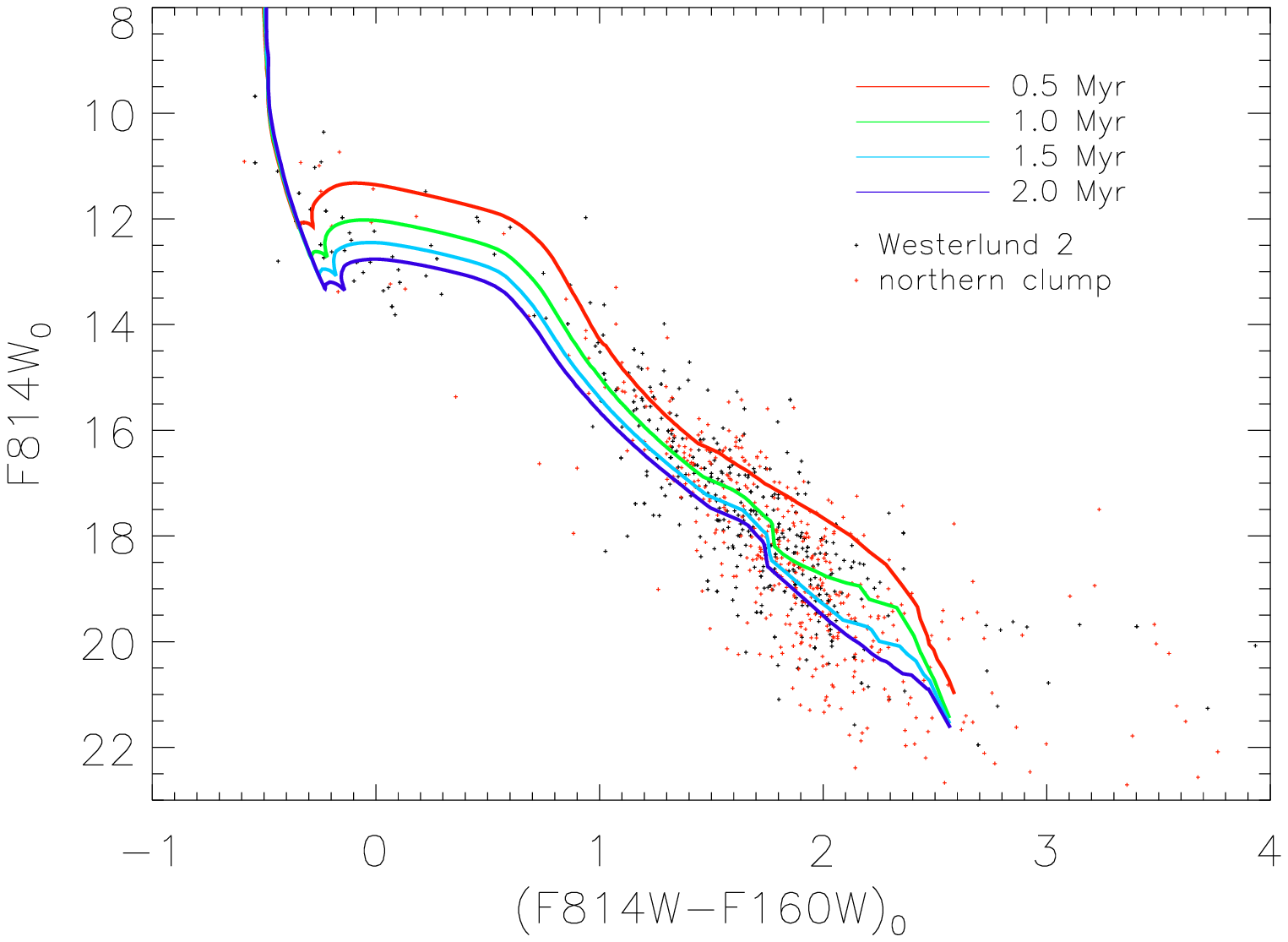}}
\caption{The $F814W$ vs. $(F814W-F160W)$ CMDs of the main body of Wd2 (black points) and the northern clump (red points). The number of objects in the main body of Wd2 was scaled to the number of objects in the northern clump (469) to facilitate the comparison. The colored lines represent the PARSEC PMS isochrones for 0.5--2.0~Myrs fitting the turn-on regions of the PMS stars towards the main sequence.}
\label{fig:scaled_CMD}
\end{figure}

As already seen in the images (see Fig. \ref{fig:reddening+I_band}) and in the spectra the main body of Wd2 contains by far more luminous ($>14$~mag) stars than the northern clump, both along the clearly visible turn-on of the PMS, as well as on the main sequence. In the northern clump the turn-on is barely visible and there are almost no stars on the main sequence up to our saturation limit of $\sim F814W=13$~mag. One reason of this apparent lack of more massive stars may well be an effect of the stochastic sampling of the mass function \citep[e.g.,][]{Baker_87} of the relatively sparsely populated northern clump. For a better comparison of the CMDs of the two clumps we reduced the number of objects in the main body of Wd2 to the same number as in the northern clump by a randomized removal of objects. In Fig. \ref{fig:scaled_CMD} one can see the such reduced CMD of the main body of Wd2 in black and the CMD of the northern clump in red. Besides the fact that the turn-on is more apparent in the main body of Wd2 there is no visible difference. The scatter of stars located on the turn-on is similar for the main body of Wd2 and the northern clump. This leads us to suggest that both clumps formed at the same time, but what we observe as the main body of Wd2 presumably formed from an initially more massive and possibly denser gas clump. The mass function (which is left for a subsequent paper) will reveal whether there is a substantial difference between the two clumps and their formation process.

We note that the lack of O3/O4 stars on the main sequence of the northern clump may imply a slightly higher age, but may also just be a consequence of stochastic sampling effects as pointed out before. In any case, for such young ages isochrones do not permit us to constrain the age more accurately, and within the accuracy of our age estimates we may consider the two clumps to be coeval.

Our CMDs show a luminosity spread among the PMS stars in the turn-on region that, if taken at face value, is compatible with an age spread of $\sim 2 $~Myr. However, considering the findings of, e.g., \citet{Baraffe_09,Hosokawa_11}, and \citet{Soderblom_14} on the impact of accretion and differences in the survival of circumstellar disks, we conclude that the adoption of $\sim 1$ Myr as cluster age is reasonable. This is also supported by the fact that the locus of the turn-on region is very sensitive to the age ($\sim 0.33$~mag~Myr$^{-1}$ at 3~Myr), which is shown by \citet{Cignoni_10ApJ} for the young cluster \object{NGC 346}.

\section{Summary and Conclusions}
\label{sec:summary}

In this paper we described the reduction of our recently obtained HST (ACS and WFC3) data of the very young star cluster Wd2. We created a photometric catalog of 17\,121 objects in six different filters, four wide-band ($F555W$, $F814W$, $F125W$, $F160W$) and two narrow-band filters ($F658N$, $F128N$).

Using the two narrow-band filters sampling the H$\alpha$ and Pa$\beta$ line, we created the first high resolution pixel-to-pixel map of the color excess $E(B-V)_g$ of the gas in Wd2. This map has a median color excess $E(B-V)_g=1.87$~mag and reveals considerable differential reddening across the region. A radial analysis of the color excess for the four quadrants (NW, NE, SW, SE) shows an almost constant increase from $\sim 1.86$~mag to values of about 2.15~mag to the Southwest and $\sim 2.0$~mag to the Southeast. To the North we have a decrease of the excess down to $\sim 1.7$~mag due to missing gas and dust. The median error due to the Poisson statistics of the electron count on the detector amounts to 0.032~mag. We estimate a contamination of the continuum in $F125W$ originating from the Pa$\beta$ line of $\sim22\%$. Following \citet{Pang_11} we infer a contamination of 10--15\% in the broad band filters due to other emission lines, which leads to an uncertainty of 0.1--0.15~mag on $E(B-V)_g$.

Using the spectroscopy catalog obtained by \citet{Alvarez_13,Rauw_07}, and \citet{Rauw_11} we translate the pixel-to-pixel gas excess map $E(B-V)_g$ into a stellar excess map $E(B-V)_\star$. The median of this map is $E(B-V)_\star = 1.55$~mag which is consistent with the mean value for the stellar excess derived with the $UBV$ TCD using the photometric catalog of \citet{Alvarez_13}.

Taking the PARSEC ZAMS \citep{Bressan_12} and fitting them to three different TCDs ($F555W-F814W$ vs. $F814W-F125W$ (see Fig. \ref{fig:F555W-F814W_F814W-F125W}), $F555W-F814W$ vs. $F125W-F160W$ (see Fig. \ref{fig:F555W-F814W_F125W-F160W}), and $(F439W-F555W)_{WFPC2}$ vs. $(F555W-F814W)_{WFPC2}$ (see Fig. \ref{fig:F439W-F555W_F555W-F814W})) allowed us to derive a value for the total-to selective extinction of $R_V=3.95 \pm 0.135$. This is in good agreement with the studies of \citet{Alvarez_13,Rauw_07,Rauw_11} and \citet{Hur_15}, who showed that Wd2 has a total to selective extinction in the visual ($R_V=A_V/E(B-V)$) of about 3.64--3.85. The value of $R_V$ lies well above the mean of the interstellar medium (ISM) of $R_V=3.1$ \citep{Cardelli_89}. This indicates that the dust grains along the line-of-sight towards Wd2 are larger than normally in the MW \citep{Mathis_81,Mathis_90,Cardelli_89}. This fact was also indicated by \citet{Alvarez_13}. This is consistent with the higher than normal slope of the $E(U-B) / E(B-V)$ TCD of $0.90 \pm 0.035$ \citep{Turner_73,Turner_76}.

In combination with the derived color excess map, the Besan\c{c}on model, and the $F814W$ vs. $(F814W-F160W)$ CMD we were able to separate the cluster stars from the foreground contamination in the region of the giant \ion{H}{2} region \object{RCW~49}. The CMD shows two well separated populations, where the bluer one consists of foreground stars representing the field population towards the Carina-Sagittarius spiral arm, while the redder one can be considered to be part of RCW~49 cloud containing the Wd2 cluster. We dereddened the selected RCW~49 population using the $E(B-V)_\star$ map which was translated into the total extinction at the pivot wavelength of the HST filters using the extinction law of \citet{Cardelli_89}. Using PARSEC isochrones we estimated a distance of $d \approx 4$~kpc, which is in good agreement with the more accurate result of \citet{Alvarez_13} of $d=4.16$~kpc, which we ultimately adopted. With this distance Wd2 is located at the same distance \citep{Gennaro_11,Bonanos_12} as Westerlund 1 \citep{Westerlund_61}, the most massive open star cluster known in our Galaxy, and is closer than NGC~3603 \citep{Harayama_08}.

The spatial density map (see Fig. \ref{fig:density_map}) of the stellar content of Wd2 reveals the splitting of Wd2 into two clumps (denoted as the main body of Wd2 and the northern clump). We infer the approximate area in which objects can be assumed to belong primarily to the respective clumps. This gave us projected radii (FWHM of the Gaussians) of $r_{\rm{Wd2}}=24.10''$ ($\sim 0.5$~pc) and $r_{\rm{nc}}=18.96''$ ($\sim 0.4$~pc) for Wd2 and the northern clump, respectively. This leads to a projected area of $\sim 0.5$~arcmin$^2$ for the main body of Wd2 and $\sim 0.3$~arcmin$^2$ for the northern clump. Including all stars with $F814W>22.75$~mag and $F160W>18.00$~mag, the mean stellar density of the main body of Wd2 is approximately $~2\,000$~stars~arcmin$^{-2}$, while the northern clump has a density of roughly $~1600$~stars~arcmin$^{-2}$. The stellar number density is therefore $\sim 20$\% smaller in the northern clump.

A comparison of the CMDs for the two clumps showed no age difference. PMS isochrones indicate for both an age between 0.5--2.0~Myr. This leads us to conclude that both clumps were born at roughly the same time ($1.0^{+1.0}_{-0.5}$~Myrs ago). Therefore, Wd2 has the same age as other very young star clusters like \object{NGC~3603} in the Carina spiral arm \citep{Pang_13}, \object{Trumpler~14} \citep{Carraro_04_Tr14} in the Carina Nebula \citep{Smith_08}, \object{R136} in the \object{LMC} \citep{Walborn_97,Sabbi_12}, \object{Arches} \citep{Figer_02,Figer_05}, and is younger than \object{Westerlund~1} \citep{Clark_05,Gennaro_11,Lim_13}. The main body of Wd2 is more massive than the northern clump and shows a well defined PMS turn-on region. The main body of Wd2 contains at least 19 O-type stars \citep{Alvarez_13,Rauw_07,Rauw_11} while the northern clump just contains one 1 B and 2 O-type stars leading to a significant stellar mass difference.

Such subclusters appear to be a common occurrence in massive star-forming regions, comprising between 2 to 20 objects \cite[see][and references therein]{Kuhn_14}. Wd2 thus shows a modest example of such a substructure.

In a forthcoming paper, we will carry out completeness tests and determine the present-day mass function as well as the mass of both clumps. \citet{Nota_15} describes the morphology of the region, the detection of a rich population of pre-main-sequence stars in the cluster, and a number of unusual protostellar objects observed in the Wd2 region.

\acknowledgements
Based on observations made with the NASA/ESA Hubble Space Telescope, obtained at the Space Telescope Science Institute, which is operated by the Association of Universities for Research in Astronomy, Inc., under NASA contract NAS 5-26555. These observations are associated with program \#13038.

E.K.G. and A.P. acknowledge support by Sonderforschungsbereich 881 (SFB 881, The Milky Way System) of the German Research Foundation, particularly via subproject B5. M.T. has been partially funded by PRIN-MIUR 2010LY5N2T.

We thank Leo Girardi for helping us to download isochrones from the Padova Trieste evolutionary model without using the web interface.

We thank the referee for the very detailed and helpful report.

{\it Facilities:} \facility{HST (STScI)}.

\bibliographystyle{apj}
\bibliography{bibliography}

\end{document}